\documentclass[12pt]{article}
\usepackage{epsfig,amsmath,amssymb,euscript}

\def\F{{\EuScript F}}

\def\S{{\cal S}}

\begin{document}

\begin{titlepage}

\begin{flushright}
CLNS~05/1911\\
MIT-CTP 3616\\
{\tt hep-ph/0504071}\\[0.2cm]
April 8, 2005
\end{flushright}

\vspace{0.2cm}
\begin{center}
\Large\bf
Theory of Charmless Inclusive {\boldmath $B$} Decays
and the Extraction of {\boldmath $V_{ub}$}
\end{center}

\vspace{0.4cm}
\begin{center}
{\sc Bj\"orn O. Lange$^a$, Matthias Neubert$^b$, and Gil Paz$^b$}\\
\vspace{0.4cm}
{\sl $^a$\,Center for Theoretical Physics\\ Massachusetts Institute of
Technology\\
Cambridge, MA 02139, U.S.A.\\[0.3cm]
$^b$\,Institute for High-Energy Phenomenology\\
Newman Laboratory for Elementary-Particle Physics, Cornell University\\
Ithaca, NY 14853, U.S.A.}
\end{center}

\vspace{0.5cm}
\begin{abstract}
\noindent 
We present ``state-of-the-art'' theoretical expressions for the triple 
differential $\bar B\to X_u\,l^-\bar\nu$ decay rate and for the 
$\bar B\to X_s\gamma$ photon spectrum, which 
incorporate all known contributions and smoothly interpolate between the 
``shape-function region'' of large hadronic energy and small invariant mass, 
and the ``OPE region'' in which all hadronic kinematical variables scale with 
$M_B$. The differential rates are given in a form which has no explicit
reference to the mass of the $b$ quark, avoiding the associated
uncertainties. Dependence on $m_b$ enters indirectly through the 
properties of the leading shape function, which can be determined by
fitting the $\bar B\to X_s\gamma$ photon spectrum. This eliminates the 
dominant theoretical uncertainties from predictions for 
$\bar B\to X_u\,l^-\bar\nu$ decay distributions, allowing for a precise 
determination of $|V_{ub}|$. In the shape-function 
region, short-distance and long-distance contributions are factorized 
at next-to-leading order in renormalization-group improved
perturbation theory. Higher-order power corrections include effects from 
subleading shape functions where they are known. When integrated over 
sufficiently large portions in phase space, our results reduce to standard
OPE expressions up to yet unknown $O(\alpha_s^2)$ terms. Predictions are 
presented for partial $\bar B\to X_u\,l^-\bar\nu$ decay rates with 
various experimental cuts.
An elaborate error analysis is performed that contains all significant 
theoretical uncertainties, including weak annihilation effects. We suggest
that the latter 
can be eliminated by imposing a cut on high leptonic invariant mass.
\end{abstract}

\end{titlepage}

\section{Introduction}

A major effort of the $B$-physics community is underway to map out the
apex of the unitarity triangle, which provides a graphical
representation of the effect of CP violation in the quark flavor sector of
the Standard Model.
One of the biggest successes of this endeavor was the precise
determination of the angle $\beta$, which has been measured with high
accuracy from the time-dependent CP asymmetry in the $B\to J/\psi K_S$ 
decay channel \cite{Aubert:2004zt,Abe:2004mz}. The length of the side
opposite the angle $\beta$ is proportional to $|V_{ub}|$. A 
high-precision determination of this quantity would enable us to test the
validity of the Standard Model and search for possible deviations 
from its predictions.

Good theoretical knowledge of strong-interaction effects in weak decays of
$B$ mesons is crucial for a reliable exploration of the flavor sector of the 
Standard Model. In particular, the determination of
the Cabibbo-Kobayashi-Maskawa (CKM) matrix elements $|V_{cb}|$ and $|V_{ub}|$
relies on an accurate description of bound-state effects in semileptonic 
decays. At present, the most precise calculations are available for inclusive 
semileptonic decays $\bar B\to X\,l^-\bar\nu$. 

The theoretical tools for the calculation of inclusive $B$ decays are
QCD factorization on the one hand
\cite{Neubert:1993ch,Neubert:1993um,Bigi:1993ex,Korchemsky:1994jb,%
Akhoury:1995fp,Bauer:2003pi,Bosch:2004th,Gardi:2004ia,Neubert:2004dd}, and 
local operator product expansions
(OPE) on the other \cite{Blok:1993va,Manohar:1993qn}. Both approaches
perform a systematic separation of long-distance hadronic quantities from
short-distance
perturbative ones, while organizing the calculation in inverse powers
of the heavy $b$-quark mass $m_b$. The OPE is an appropriate
tool for the calculation of total inclusive rates 
(for example in $\bar B\to X_c\,l^-\bar\nu$ decay) or for
partial rates integrated over sufficiently large regions in phase space, 
where all components of the final-state hadronic momentum $P_X^\mu$ 
are large compared 
to $\Lambda_{\rm QCD}$. QCD factorization, on the other hand, is better 
suited for the calculation of partial rates and spectra near kinematical 
boundaries, where typically some components of $P_X^\mu$ are large, while 
the invariant hadronic mass $M_X=\sqrt{P_X^2}$ is small. 
For example, any $\bar B\to X_u\,l^-\bar\nu$ event can be described 
with three independent kinematical variables, a useful choice of which is 
\cite{Bosch:2004th,Bosch:2004bt}
\begin{equation}
   P_l = M_B - 2 E_l \,, \qquad 
   P_- = E_X + |\vec P_X| \,, \qquad 
   P_+ = E_X - |\vec P_X| \,.
\end{equation}
Here $P_\pm$ are the light-cone components of the hadronic final-state 
momentum along the jet direction, $E_l$ is the charged-lepton energy, $E_X$ 
is the jet energy, and $\vec P_X$ is the jet momentum, all measured 
in the $B$-meson rest frame. The phase space for these variables is
\begin{equation}\label{eq:simplePhaseSpace}
   \frac{M_\pi^2}{P_-} \le P_+ \le P_l \le P_- \le M_B \,,
\end{equation}
with $M_\pi$ being the mass of the lightest possible hadronic final state. 
The product $P_+ P_- = M_X^2$ is the hadronic invariant mass squared. In 
order to avoid large backgrounds from $b\to c$ transitions, all measurements 
of $|V_{ub}|$ are in one way or another restricted
to the region of phase space where $P_+ P_- < M_D^2$. If the quantity
$P_-$ takes values near its maximum at $M_B$, then
$P_+$ is restricted to a region of order $M_D^2/M_B$, which is
numerically comparable to $\Lambda_{\rm QCD}$. This means that there
are three parametrically different energy scales in the problem: the
mass $M_B$ of the initial state, the mass of the final hadronic state
$\sim \sqrt{M_B\Lambda_{\rm QCD}}$, and the low scale $\Lambda_{\rm QCD}$ 
at which perturbation theory breaks down and hadronic physics must be 
parameterized in terms of non-perturbative matrix elements. QCD factorization 
disentangles the effects from these scales, so that the perturbative
contributions can be expanded in powers of $\alpha_s(\mu_h)$ with 
$\mu_h \sim m_b$ (giving rise to ``hard functions'') and $\alpha_s(\mu_i)$ 
with $\mu_i \sim\sqrt{m_b\Lambda_{\rm QCD}}$ (giving rise to ``jet 
functions''). 

It is important to note that the heavy-quark expansions
valid in these two kinematical regions are not identical, because the power 
counting rules differ in the two regimes. 
Also the nature of the non-perturbative inputs is different. 
In the OPE region, non-perturbative
physics is encoded in a few hadronic parameters, and the heavy-quark 
expansion is the usual Wilsonian expansion in local operators. In the endpoint
(or shape-function) region, the presence of multiple scales complicates the
power counting, and the interplay between soft and collinear modes gives
rise to large non-localities. As a result,
non-perturbative physics is described by hadronic structure functions 
called ``shape functions'', and the heavy-quark expansion is an expansion in 
non-local string operators defined on the light-cone. The connections 
between the two regimes is that {\em moments\/} of the shape functions can
be expressed in terms of local operators.

The goal of the present work is to develop a formalism that smoothly 
interpolates between the two kinematical regimes (see \cite{Tackmann:2005ub}
for a related discussion, which is however restricted to the tree
approximation). This is essential for
building an event generator for inclusive $\bar B\to X_u\,l^-\bar\nu$ and
$\bar B\to X_s\gamma$ decays, which can be used to study 
partial and differential decay rates in different kinematical domains. 
In the shape-function region, our approach relies on exact QCD factorization 
theorems, which exist in every order of power counting. They allow
us to systematically disentangle short- and long-distance physics and,
in the process, resum parametrically large logarithms order by order in 
perturbation theory. This factorization can be done with high accuracy for
the terms of leading power in $1/m_b$, and with somewhat less sophistication
for the first-order power corrections. For the second-order power corrections,
we only include contributions that do not vanish when integrated
over all phase space. This is a safe approximation; the effects of the 
remaining $1/m_b^2$ terms can to a large extent 
be absorbed by a redefinition of the subleading shape functions arising at
order $1/m_b$.

Our formalism is ``optimized'' for the shape-function region in the
sense that sophisticated theoretical technology is applied in
this regime. However, when our expressions for the 
differential decay rates are integrated over sufficiently wide domains, 
they automatically reduce to the simpler results that can be derived using
the OPE approach, up to yet unknown terms of $O(\alpha_s^2)$. 
The moment relations for the shape functions are crucial
in this context. Note that local $1/m_b^2$ corrections in the OPE receive
contributions from terms of leading power ($1/m_b^0$), subleading power 
($1/m_b$), and sub-subleading power ($1/m_b^2$) in the shape-function region, 
so the transition is highly non-trivial.  
In implementing the program outlined here, we include 
all presently known information on the triple differential 
$\bar B\to X_u\,l^-\bar\nu$ decay rate and on the differential 
$\bar B\to X_s\gamma$ decay rate in a single, unified framework. 
We neglect, for simplicity, hadronic power corrections of order $1/m_b^3$ and
higher, which are known to have a negligible effect on the observables 
considered here. The only possible exception is contributions from ``weak
annihilation'', which are estimated as part of our error analysis. 
We also ignore the existing results on $O(\beta_0\alpha_s^2)$ radiative
corrections for some single-differential distributions, because the 
corresponding corrections are not known for the double or triple differential 
$\bar B\to X_u\,l^-\bar\nu$ decay spectra. 
While these $O(\beta_0\alpha_s^2)$ terms are sometimes found to be large
when naive perturbation theory in $\alpha_s(m_b)$ is used, their effects
are expected to be small in our scheme, which is based on a complete scale
separation using QCD factorization. We see no reason why the 
$\beta_0\alpha_s^2$ terms should be
enhanced compared to other, unknown corrections of $O(\alpha_s^2)$.

A technical complication in realizing the approach described here has to do 
with the treatment of phase-space factors. The heavy-quark expansion of 
the hadronic tensor for $\bar B\to X_u\,l^-\bar\nu$ decay gives rise to
expressions that are singular at certain points in phase space. One way to 
avoid these singularities is to also expand phase-space factors order by 
order in $1/m_b$ (see, e.g., the treatment in \cite{Bosch:2004cb}). However, 
since this expansion depends on the kinematical cuts of any given analysis, 
it cannot be implemented in a straightforward way in an event generator. An
alternative is to reorganize the heavy-quark expansion in such a way that 
the expansion parameter is related to {\em hadronic\/} (as opposed to 
partonic) kinematical variables, in which case kinematical singularities are
always canceled by exact phase-space factors. Following this strategy, we
obtain expressions for decay distributions and partial decay rates which
are free of explicit reference to partonic quantities such as the
$b$-quark mass. A dependence on $m_b$ enters only implicitly via the 
first moment of the leading-order shape function 
$\hat S(\hat \omega)$. The philosophy of our approach is that this 
function\footnote{More precisely, we define a new shape function 
$\hat\S(\hat\omega)$ by the combination of leading and subleading shape 
functions contributing to $\bar B\to X_s\gamma$ decay, and we will use the 
same function to make predictions for $\bar B\to X_u\,l^-\bar\nu$ decay 
distributions.}
is extracted experimentally from a fit to the $\bar B\to X_s\gamma$ photon 
spectrum, which has been measured with good precision in the region 
where $P_+ = M_B - 2 E_\gamma \sim \Lambda_{\rm QCD}$. This is analogous 
to the extraction of parton distribution functions from deep inelastic 
scattering. The
photon spectrum is experimentally accessible to energies as low as
1.8\,GeV, which corresponds to a sampling of the shape function for
values of $\hat\omega$ up to about 1.7\,GeV. Once the shape function has 
been extracted over this range, we can use it to obtain predictions for 
arbitrary partial $\bar B\to X_u\,l^-\bar\nu$ decay rates with cuts. In doing 
so, the residual hadronic uncertainties in the extraction of $|V_{ub}|$ only 
enter at the level of power corrections.

We emphasize that the program outlined above is equivalent to an approach put 
forward in \cite{Neubert:1993um} and later refined in 
\cite{Leibovich:1999xf,Leibovich:2000ey,Neubert:2001sk}, in which $|V_{ub}|$ 
is extracted with the help of shape-function independent relations between 
weighted integrals over differential decay distributions in 
$\bar B\to X_s\gamma$ and $\bar B\to X_u\,l^-\bar\nu$. The experimental 
error in the results for these weighted integrals corresponds, in our 
approach, to the error in the prediction of $\bar B\to X_u\,l^-\bar\nu$ 
partial rates resulting from the experimental uncertainty in the extraction of 
the shape function from the $\bar B\to X_s\gamma$ photon spectrum. While the 
shape-function independent relations are very elegant, it is more convenient 
for the construction of a generator to have a formulation where the shape 
function is used as an input. In this way, it is possible to impose arbitrary 
cuts on kinematical variables without having to recompute the weight 
functions in each case.

The paper is structured as follows: In Section~\ref{sec:BsGamma} we collect 
the relevant formulae for the calculation of the $\bar B\to X_s\gamma$ 
photon spectrum. These expressions can be used to extract
the leading non-perturbative structure function from experiment. An
analogous presentation for the triple differential decay rate in 
$\bar B\to X_u\,l^-\bar\nu$ decays is presented in Section~\ref{sec:BuLnu}. In 
order to perform a numerical analysis one needs to rely on parameterizations 
of the shape functions. A collection of several useful functional forms is 
given in Section~\ref{sec:strucfunc}. In Section~\ref{sec:analysio} we present
a full error analysis of partial $\bar B\to X_u\,l^-\bar\nu$ decay rates for a 
variety of experimental cuts. We also explore the sensitivity of the results 
to the $b$-quark mass and to the functional forms adopted for the shape 
functions. Section~\ref{sec:concl} contains our conclusions.

\section{Inclusive radiative decays}
\label{sec:BsGamma}

The decay process $\bar B\to X_s \gamma$, while more complex in its
short-distance physics, is considerably simpler in its kinematics than the
semileptonic process $\bar B\to X_u\,l^-\bar\nu$. Since the
radiated photon is on-shell, the hadronic variables $P_\pm$ that
describe the momentum of the $X_s$ system are trivially related to the
photon energy $E_\gamma$ by $P_+ = M_B - 2 E_\gamma$ and 
$P_-=M_B$. In the crudest approximation, namely at tree level and leading
power, the photon-energy spectrum is directly proportional to the
leading shape function, $d\Gamma_s/dE_\gamma \propto \hat S(P_+)$. 
In this section we collect all relevant formulae needed to compute the 
$\bar B\to X_s\gamma$ photon spectrum or, equivalently, the invariant 
hadronic mass distribution. It is implicitly assumed that these spectra are 
sufficiently ``smeared'' (e.g., by experimental resolution)
to wash out any sharp hadronic structures. In cases where the 
resolution is such that the $K^{*}$ resonance peak is observed, it 
can be accounted for by combining the formulae 
in this section with the prescription for subtracting the $K^*$ peak 
proposed in \cite{Kagan:1998ym}.

The differential $\bar B\to X_s\gamma$ decay rate can be written as
\begin{equation}\label{BsG:general}
   \frac{d\Gamma_s}{d E_\gamma} = \frac{G_F^2\alpha}{2\pi^4}\,E_\gamma^3\,
   |V_{tb} V_{ts}^*|^2\,\overline{m}_b^2(\mu_h)\,
   [C_{7\gamma}^{\rm eff}(\mu_h)]^2\,U(\mu_h,\mu_i)\,\F_\gamma(P_+) \,,
\end{equation}
where the structure function $\F_\gamma$ depends on the photon energy
via $P_+=M_B-2E_\gamma$. The prefactor contains the
electromagnetic fine-structure constant $\alpha$ normalized at $q^2=0$, 
two powers of the running $b$-quark mass (defined in the $\overline{\rm MS}$ 
scheme) originating from the electromagnetic dipole
operator $Q_{7\gamma}$ in the effective weak 
Hamiltonian, and the square of the corresponding Wilson coefficient 
$C_{7\gamma}^{\rm eff}$, which is needed at next-to-leading order in 
renormalization-group improved perturbation theory \cite{Chetyrkin:1996vx}.
Renormalization-group running from the hard scale 
$\mu_h\sim m_b$ to the intermediate scale 
$\mu_i\sim\sqrt{m_b\Lambda_{\rm QCD}}$ gives rise to the evolution factor
$U(\mu_h,\mu_i)$, whose explicit form is discussed in 
Appendix~\ref{apx:Sudakovs}.
We keep $U$ and $(C_{7\gamma}^{\rm eff})^2$ outside of
the structure function $\F_\gamma$; it is understood that when
combining the various terms in (\ref{BsG:general}) all perturbative quantities
should be expanded for consistency to the required order in $\alpha_s$. 

\subsection{Leading-power factorization formula}

At leading order in $1/m_b$ the structure
function $\F_\gamma$ factorizes as \cite{Neubert:2004dd}
\begin{equation}\label{BsG:LO}
   \F_\gamma^{\rm (0)}(P_+) = |H_s(\mu_h)|^2 \int_0^{P_+}\!d\hat\omega\,
   m_b\,J(m_b(P_+ - \hat\omega),\mu_i)\,\hat S(\hat\omega,\mu_i) \,. 
\end{equation}
At this order a single non-perturbative parton distribution function arises, 
called the leading shape function \cite{Neubert:1993um} and denoted by 
$\hat S(\hat \omega,\mu_i)$. Our notation is adopted from 
\cite{Bosch:2004th,Bosch:2004cb}: hatted shape functions have support for 
$\hat\omega\ge 0$. The function $\hat S$ is
defined in terms of a non-local matrix element in heavy-quark effective
theory (HQET). Renormalization-group
running between the intermediate scale and a low hadronic scale is
avoided when using the shape functions renormalized at the intermediate 
scale $\mu_i$. Evolution effects below this scale are universal (i.e., 
process independent) and so can be absorbed into the renormalized shape 
function. Short-distance contributions from scales above $\mu_h\sim m_b$ are 
included in the hard function $H_s$, which in practice is obtained
by matching the effective weak Hamiltonian onto a current operator in 
soft-collinear effective theory (SCET). 
At next-to-leading order in perturbation theory, the result reads
\begin{eqnarray}\label{Hsresult}
   H_s(\mu_h)
   &=& 1 + \frac{C_F\alpha_s(\mu_h)}{4\pi} \left( -2\ln^2\frac{m_b}{\mu_h}
    + 7\ln\frac{m_b}{\mu_h} - 6 - \frac{\pi^2}{12} \right)
    + \varepsilon_{\rm ew} \nonumber\\
   &&\mbox{}+ \frac{C_{8g}^{\rm eff}(\mu_h)}{C_{7\gamma}^{\rm eff}(\mu_h)}\,
    \frac{C_F\alpha_s(\mu_h)}{4\pi}
    \left( - \frac83 \ln\frac{m_b}{\mu_h} + \frac{11}{3} - \frac{2\pi^2}{9}
    + \frac{2\pi i}{3} \right) \nonumber\\
   &&\mbox{}+ \frac{C_1(\mu_h)}{C_{7\gamma}^{\rm eff}(\mu_h) }\,
    \frac{C_F\alpha_s(\mu_h)}{4\pi}
    \left( \frac{104}{27} \ln\frac{m_b}{\mu_h} + g(z)
    - \frac{V_{ub} V_{us}^*}{V_{tb} V_{ts}^*}\,\big[ g(0) - g(z) \big] \right)
    + \varepsilon_{\rm peng} \,,
\end{eqnarray}
where the variable $z=(m_c/m_b)^2$ denotes the ratio of quark
masses relevant to charm-loop penguin diagrams, and the ``penguin function'' 
$g(z)$ can be approximated by the first few terms of its Taylor expansion,
\begin{eqnarray}
   g(z) &=& - \frac{833}{162} - \frac{20\pi i}{27} + \frac{8\pi^2}{9}\,z^{3/2}
    \nonumber\\
   &&\mbox{}+ \frac{2z}{9} \left[ 48 - 5\pi^2\! - 36\zeta_3
    + (30\pi - 2\pi^3) i + \!(36 - 9\pi^2 + 6\pi i) \ln z
    + \!(3 + 6\pi i) \ln^2 z + \ln^3 z \right] \nonumber\\
   &&\mbox{}+ \frac{2z^2}{9} \left[ 18 + 2\pi^2 - 2\pi^3 i
    + (12 - 6\pi^2) \ln z + 6\pi i\ln^2 z + \ln^3 z \right] \nonumber\\
   &&\mbox{}+ \frac{z^3}{27} \left[ -9 - 14\pi^2 + 112\pi i
    + (182 - 48\pi i) \ln z - 126\ln^2 z \right] + \dots \,.
\end{eqnarray}
The Wilson coefficients $C_1$ and $C_{8g}^{\rm eff}$ in (\ref{Hsresult}) 
multiply the current-current operators $Q_1^{u,c}$ and the chromo-magnetic 
dipole operator $Q_{8g}$ in the effective weak Hamiltonian.
The quantities $\varepsilon_{\rm ew}\approx -1.5\%$ and 
$\varepsilon_{\rm peng}\approx -0.6\%$ 
account for small electroweak corrections 
and the effects of penguin contractions of operators other than $Q_1^{u,c}$,
respectively. 
The differential decay rate (\ref{BsG:general}) is formally
independent of the matching scales $\mu_h$ and $\mu_i$. The $\mu_h$ 
dependence of the evolution factor $U(\mu_h,\mu_i)$ cancels the scale 
dependence of the product 
$\overline{m}_b^2(\mu_h)\,[C_{7\gamma}^{\rm eff}(\mu_h)]^2\,|H_s(\mu_h)|^2$, 
while its $\mu_i$ dependence compensates the scale dependence of the
convolution integral $J(\mu_i)\otimes \hat S(\mu_i)$. 

Finally let us discuss the jet function $J$, which appears as the 
hard-scattering kernel in the convolution integral in (\ref{BsG:LO}). It can 
be written in terms of distributions that act on the shape function $\hat S$. 
At one-loop order, the jet function is given by 
\cite{Bauer:2003pi,Bosch:2004th}
\begin{equation}\label{BsG:jetfunc}
   J(p^2,\mu) = \delta(p^2) \left[ 1
   + \frac{C_F \alpha_s(\mu)}{4\pi}(7-\pi^2) \right]
   + \frac{C_F \alpha_s(\mu)}{4\pi} \left[ \frac{1}{p^2}
   \left( 4\ln \frac{p^2}{\mu^2} - 3\right) \right]_*^{[\mu^2]} \,,
\end{equation}
where the star distributions have the following effect on a function
$f$ when integrated over a domain $Q^2$ \cite{DeFazio:1999sv}:
\begin{eqnarray}\label{stardistris}
   \int_{\le 0}^{Q^2} dp^2\,\left[\frac{1}{p^2} \right]_*^{[\mu^2]} f(p^2)
   &=& \int_0^{Q^2} dp^2\,\frac{f(p^2)-f(0)}{p^2} \,
    + f(0) \,\ln \frac{Q^2}{\mu^2} \,, \nonumber\\
   \int_{\le 0}^{Q^2} dp^2\,\left[\frac{1}{p^2} \ln \frac{p^2}{\mu^2}
    \right]_*^{[\mu^2]} f(p^2) 
   &=& \int_0^{Q^2} dp^2\,\frac{f(p^2)-f(0)}{p^2}\ln \frac{p^2}{\mu^2}
    + \frac{f(0)}{2}\,\ln^2 \frac{Q^2}{\mu^2} \,. 
\end{eqnarray}

\subsection{Kinematical power corrections}

There exists a class of power corrections to (\ref{BsG:LO}) that do not
involve new hadronic quantities. Instead, the power suppression results from
the restriction of certain variables ($P_+$ in the present case) 
to a region where they are kinematically suppressed (here $P_+\ll M_B$).
The corresponding terms are known in fixed-order perturbation theory, without 
scale separation and renormalization-group resummation 
\cite{Greub:1996tg,Ali:1995bi} (see also 
\cite{Kagan:1998ym}). To perform a complete RG analysis of even the 
first-order terms in $1/m_b$ is beyond the scope of the present work. 
Since, as we will see later, power corrections only account for small
corrections to the decay rates, an approximate treatment will suffice. 
To motivate it, we note the following two facts \cite{Neubert:2004dd}: 
First, while the 
anomalous dimensions of the relevant subleading SCET and HQET operators are 
only known for a few cases \cite{Hill:2004if}, the leading Sudakov double 
logarithms are the 
same as for the terms of leading power, because they have a geometric origin in
terms of Wilson lines \cite{Becher:2003kh}. The leading 
Sudakov double logarithms are therefore the same as those resummed into the 
function $U$ in (\ref{BsG:general}). Secondly, the kinematical power
corrections in $\bar B\to X_s\gamma$ decay  
are associated with gluon 
emission into the hadronic final state $X_s$. Because of the kinematical 
restriction to low-mass final states, i.e.\ $M_X^2\sim M_B\Lambda_{\rm QCD}$, 
we associate a coupling $\alpha_s(\bar\mu)$ with these terms, where typically
$\bar\mu\sim\mu_i$. Strictly speaking, however, the scale ambiguity associated
with the choice of $\bar\mu$ could only be resolved by computing the relevant
anomalous dimensions. 

Within this approximation, the kinematical power corrections to the structure
function $\F_\gamma$ can be extracted from \cite{Neubert:2004dd,Kagan:1998ym}. 
We find it convenient to express the result in terms of the variable
\begin{equation}\label{xdef}
   x = \frac{P_+ - \hat\omega}{M_B - P_+} \,,
\end{equation}
which in the shape-function region scales like $\Lambda_{\rm QCD}/m_b$.
We obtain
\begin{eqnarray}\label{realrads}
   \F_\gamma^{\rm kin}(P_+)
   &=& \frac{1}{M_B-P_+}\,\frac{C_F\alpha_s(\bar\mu)}{4\pi} \hspace{-0.1cm}
    \sum_{\scriptsize\begin{array}{c} i,j=1,7,8 \\[-0.1cm] i\le j \end{array}}
    \hspace{-0.1cm}
    \frac{C_i(\mu_h)\,C_j(\mu_h)}{C_{7\gamma}^{\rm eff}(\mu_h)^2}
    \int_0^{P_+}\!d\hat\omega\,\hat S(\hat\omega,\mu_i)\,h_{ij}(x)
    \nonumber\\[-0.3cm]
   &&\mbox{}- \frac{\lambda_2}{9m_c^2}\,
    \frac{C_1(\mu_h)}{C_{7\gamma}^{\rm eff}(\mu_h)}\,\hat S(P_+,\mu_i) \,.
\end{eqnarray}
The coefficient functions $h_{ij}(x)$ are 
\begin{eqnarray}
   h_{77}(x) &=& -3(5+2x) + 2(8+9x+3x^2) \ln\left( 1 + \frac{1}{x} \right) ,
    \nonumber\\
   h_{88}(x) &=& \frac29\,(1+3x+4x^2+2x^3) \left[ 2\ln\frac{m_b}{m_s}
    - \ln\left( 1 + \frac{1}{x} \right) \right]
    - \frac19\,(3+9x+16x^2+8x^3) \,, \nonumber\\
   h_{78}(x) &=& \frac23\,(5+8x+4x^2) - \frac83\,x(1+x)^2\,
    \ln\left( 1 + \frac{1}{x} \right) , \nonumber\\
   h_{11}(x) &=& \frac{16}{9} \int_0^1\!du\,(1+x-u) 
    \left|\,\frac{z(1+x)}{u}\,
    G\!\left(\frac{u}{z(1+x)}\right) + \frac12\,\right|^2 , \nonumber\\
   h_{17}(x) &=& -3 h_{18}(x)
    = - \frac83 \int_0^1\!du\,u\,\mbox{Re}\left[\,
    \frac{z(1+x)}{u}\,G\!\left(\frac{u}{z(1+x)}\right) + \frac12 \,\right] , 
\end{eqnarray}
where as before $z=(m_c/m_b)^2$, and
\begin{equation}
   G(t) = \left\{ \begin{array}{ll}
    -2\arctan^2\!\sqrt{t/(4-t)} & ~;~ t<4 \,, \\[0.1cm]
    2 \left( \ln\!\Big[(\sqrt{t}+\sqrt{t-4})/2\Big]
    - \displaystyle\frac{i\pi}{2} \right)^2 & ~;~ t\ge 4 \,.
   \end{array} \right.
\end{equation}
In the shape-function region the expressions for $h_{ij}(x)$ could, if 
desired, be expanded in a power series in $x=O(\Lambda_{\rm QCD}/m_b)$, 
and this would generate a series of power-suppressed terms 
$\F_\gamma^{{\rm kin}(n)}(P_+)$ with $n\ge 1$, where the superscript ``$n$'' 
indicates the order in the $1/m_b$ expansion. Note that this expansion would
contain single logarithms $\ln x\sim\ln(\Lambda_{\rm QCD}/m_b)$. These are
precisely the logarithms that would be resummed in a more proper 
treatment using effective field-theory methods.

Outside the shape-function region the variable $x$ can take on arbitrarily
large positive values, and $\F_\gamma^{\rm kin}(P_+)$ is no longer power 
suppressed. Note that for $P_+\to M_B$ (corresponding to $x\to\infty$ and 
$E_\gamma\to 0$) most functions $h_{ij}(x)$ grow like $x^2$ or weaker, so 
that the spectrum tends to a constant. The only (well known) exception is 
$h_{88}(x)$, which grows like $x^3$, giving rise to a $1/E_\gamma$ 
soft-photon singularity \cite{Ali:1995bi}.
The main effect of the kinematical power corrections (\ref{realrads}) to the
photon spectrum is to add a radiative tail extending into the
region of small photon energies. These corrections therefore become the more 
significant the larger the integration domain over $E_\gamma$ is.

\subsection{Subleading shape-function contributions}
\label{subsec:BsG:SSF}

At order $1/m_b$ in power counting, different combinations of 
subleading shape functions enter the $\bar B\to X_s\gamma$ and 
$\bar B\to X_u\,l^-\bar\nu$ decay distributions 
\cite{Bauer:2001mh,Bauer:2002yu,Leibovich:2002ys,Neubert:2002yx}.
They provide the dominant hadronic power corrections, which must be 
combined with 
the kinematical power corrections discussed in the previous section.
We include their effects using the results of recent calculations in
\cite{Bosch:2004cb,Lee:2004ja,Beneke:2004in}. Little is known about the 
subleading shape functions apart from expressions for their first few
moments. In particular, the norms of these functions vanish at tree level, 
while their first moments are determined by the HQET parameters $\lambda_1$ 
and $\lambda_2$, which are defined via the forward $B$-meson matrix elements 
of the kinetic-energy and the chromo-magnetic operators, respectively 
\cite{Falk:1992wt}. 

For the case of $\bar B\to X_s\gamma$ decay, 
subleading shape-function contributions are currently only known for 
the matrix elements of the dipole 
operator $Q_{7\gamma}$, and the corresponding hard and jet functions
have been computed at tree level. Adopting the notations of 
\cite{Bosch:2004cb}, the relevant subleading shape functions are 
$\hat t(\hat\omega)$, $\hat u(\hat\omega)$, and $\hat v(\hat\omega)$. 
An additional function, called $s_0$, has been absorbed by a redefinition of 
the leading shape function, and it is included in our definition of 
$\hat S(\hat\omega)$. 
Roughly speaking, $\hat u(\hat\omega)$ is the ``light-cone
generalization'' of the local HQET kinetic-energy operator. The
functions $\hat v(\hat\omega)$ and $\hat t(\hat\omega)$ are both 
generalizations of the local chromo-magnetic HQET operator, but 
$\hat t(\hat\omega)$ contains also a
light-cone chromo-electric operator, which has no equivalent in the
local OPE expansion. (Such a contribution arises since there are 
two external 4-vectors in the SCET expansion, $n$ and $v$, while there is 
only $v$ in the HQET expansion.) The contribution of subleading shape
functions to the $\bar B\to X_s\gamma$ photon spectrum is
\begin{equation}\label{BsgSSF}
   \F_\gamma^{\rm hadr(1)}(P_+) = \frac{1}{M_B-P_+}
   \left[ - (\bar\Lambda - P_+)\,\hat S(P_+) - \hat t(P_+)
   + \hat u(P_+) - \hat v(P_+) \right] . 
\end{equation}
Compared to \cite{Bosch:2004cb}, we have replaced $1/m_b$ with 
$1/(M_B-P_+)$ in the prefactor, which is legitimate at this order. 
(The form of the shape 
functions restricts $P_+$ to be of order $\Lambda_{\rm QCD}$.)
The appearance of the HQET parameter $\bar\Lambda=(M_B-m_b)_{m_b\to\infty}$ 
is peculiar to the 
subleading shape-function contributions. This quantity is defined via the
first moment of the leading-order shape function \cite{Bosch:2004th}.

The formula given above can be modified to suit the purpose of
extracting the shape function from the photon spectrum better. To this
end, we absorb a linear combination of the subleading shape functions into
a redefinition of the leading shape function, in such a way that the
moment relations for this function remain unchanged to the
order we are working. This is accomplished by defining
\begin{equation}\label{def:ScriptS}
   \hat\S(\hat \omega)\equiv \hat S(\hat\omega)
   + \frac{2(\bar\Lambda-\hat\omega)\,\hat S(\hat\omega) - \hat t(\hat\omega)
           + \hat u(\hat\omega) - \hat v(\hat\omega)}{m_b} \,.
\end{equation}
When using $\hat\S$ instead of $\hat S$ in the leading-power formula 
(\ref{BsG:LO}), the subleading shape-function contribution becomes
\begin{equation}\label{BsgSSF2}
   \F_\gamma^{\rm hadr(1)}(P_+) = - \frac{3(\bar\Lambda-P_+)}{M_B-P_+}\,
   \hat\S(P_+) \,. 
\end{equation}

The hatted shape functions used in the present work are related to the 
original definitions in \cite{Bosch:2004cb} by 
\begin{eqnarray}
   \hat S(\hat\omega) &=& S(\bar\Lambda-\hat\omega)
    + \frac{s_0(\bar\Lambda-\hat\omega)}{m_b} \,, \nonumber\\
   \hat t(\hat\omega) &=& t(\bar\Lambda-\hat\omega) \,, \qquad 
    \hat u(\hat\omega) = u(\bar\Lambda-\hat\omega) \,, \qquad
    \hat v(\hat\omega) = v(\bar\Lambda-\hat\omega) \,,
\end{eqnarray}
where the unhatted functions have support on the interval between $-\infty$ 
and $\bar\Lambda$. It is convenient to rewrite
$\bar\Lambda-\hat\omega=\omega+\Delta\omega$, where
\begin{equation}\label{Delw}
   \Delta\omega\equiv \bar\Lambda - (M_B - m_b) 
   = \frac{\lambda_1+3\lambda_2}{2m_b} + \dots
\end{equation}
accounts for the mismatch between the HQET parameter $\bar\Lambda$ and the
difference $(M_B-m_b)$ due to power-suppressed terms in the $1/m_b$ 
expansion \cite{Neubert:1993mb}. It follows that the variable 
$\omega=(M_B-m_b)-\hat\omega$ runs from $-\infty$ to 
$(M_B-m_b)$. The moment relations for the leading and subleading shape 
functions derived in \cite{Neubert:1993um} and 
\cite{Bosch:2004cb,Bauer:2001mh} can be summarized as
\begin{eqnarray}\label{SSF:moments}
   \hat S(\hat\omega)
   &\equiv& S(\omega+\Delta\omega) + \frac{s_0(\omega+\Delta\omega)}{m_b}
    = \delta(\omega) - \frac{\lambda_1}{6}\,\delta''(\omega)
    + \frac{\lambda_1+3\lambda_2}{2m_b}\,\delta'(\omega) + \dots \,,
    \nonumber\\
   \hat t(\hat\omega)
   &\equiv& t(\omega+\Delta\omega) = \lambda_2\,\delta'(\omega) + \dots \,, 
    \nonumber\\
   \hat u(\hat\omega) 
   &\equiv& u(\omega+\Delta\omega) = - \frac{2\lambda_1}{3}\,\delta'(\omega)
    + \dots \,, \nonumber\\
   \hat v(\hat\omega) 
   &\equiv& v(\omega+\Delta\omega) = - \lambda_2\,\delta'(\omega) + \dots \,.
\end{eqnarray}
The function $\hat\S$ has the same moment expansion as $\hat S$.
The hadronic parameter $\lambda_2$ determines the leading contribution to 
the hyperfine splitting between the masses of $B$ and $B^*$ mesons through 
$m_{B^*}^2-m_B^2=4\lambda_2+O(1/m_b)$ \cite{Falk:1992wt}, from which it
follows that $\lambda_2\approx 0.12$\,GeV$^2$. The value of the parameter
$\lambda_1$ is more uncertain. In much the same way as the $b$-quark pole 
mass, it is affected by infrared renormalon ambiguities 
\cite{Martinelli:1995zw,Neubert:1996zy}.
It is therefore better to eliminate $\lambda_1$ in favor of some 
observable, for which we will choose the width of the
leading shape function. 

\subsection{Residual hadronic power corrections}
\label{subsec:BsGHQE}

At order $1/m_b^2$ a new set of sub-subleading shape
functions enter, which so far have not been classified completely in the 
literature. Since the functional form of even the subleading shape
functions is rather uncertain, there is no need
to worry too much about the precise form of sub-subleading shape functions.
Most of their effects can be absorbed into the subleading functions. An
exception, however, are terms that survive when the sub-subleading shape 
functions are integrated over a wide domain. Whereas the norms of all 
subleading ($\sim 1/m_b$) shape functions vanish, the norms of the 
sub-subleading shape functions ($\sim 1/m_b^2$) are in general 
non-zero and given in terms of the heavy-quark parameters $\lambda_1$ and 
$\lambda_2$. (At tree level, the class of functions with non-zero norm
has been studied in \cite{Tackmann:2005ub}.) Our strategy in the present work
will be as follows: We start from the well-known expressions for the 
(tree-level) second-order power corrections to the $\bar B\to X_s\gamma$ 
photon spectrum \cite{Falk:1993dh} (and similarly for the triple-differential 
$\bar B\to X_u\,l^-\bar\nu$ decay distribution 
\cite{Blok:1993va,Manohar:1993qn}, see Section~\ref{sec:MWterms}). They are of 
the form $\lambda_i/m_b^2$ times one of the singular distributions 
$\delta(p^2)$, $\delta'(p^2)$, or $\delta''(p^2)$, where 
$p^2=(m_b v-q)^2$ is the invariant partonic mass squared of the final-state 
jet. As mentioned earlier, the power counting in the shape-function region is 
different from the one used in OPE calculations, and indeed a good portion
of the $1/m_b^2$ terms in the OPE is already accounted for by the contributions
proportional to the leading and subleading shape functions in (\ref{BsG:LO}) 
and (\ref{BsgSSF}). We identify the corresponding terms
using the moment relations for the shape functions in (\ref{SSF:moments}).
In particular, this reproduces all terms at order $1/m_b^2$ in the OPE 
which contain derivatives of $\delta(p^2)$. We include the remaining terms
of the form $(\lambda_i/m_b^2)\,\delta(p^2)$ by replacing
\begin{eqnarray}
   \delta(p^2)
   &=& \delta(p_+ p_-)
    = \frac{1}{p_- - p_+}  \int d\omega\,\delta(p_+ + \omega)\,\delta(\omega)
    \nonumber\\
   &\to& \frac{1}{P_- - P_+} \int d\hat\omega\,\delta(P_+ - \hat\omega)\,
    \hat S(\hat\omega) = \frac{\hat S(P_+)}{P_- - P_+} \,.
\end{eqnarray}
Here $p_\pm$ are the light-cone projections of the partonic momentum $p^\mu$, 
which are related to the hadronic quantities $P_\pm$ by 
$P_\pm=p_\pm+(M_B-m_b)$. Similarly, $\hat\omega=(M_B-m_b)-\omega$.

The result of these manipulations is
\begin{equation}
   \F_\gamma^{\rm hadr(2)}
   = \frac{\lambda_1}{(M_B-P_+)^2}\,\hat S(P_+) \,.
\end{equation}
Together with (\ref{BsG:LO}) and (\ref{BsgSSF}) this accounts for
all known first- and second-order power corrections to the 
$\bar B\to X_s\gamma$ photon spectrum, both in the shape-function region and 
in the OPE region. The redefinition (\ref{def:ScriptS}) of the
leading shape function from $\hat S$ to $\hat\S$ leaves the form of the 
second-order power corrections unaffected. 

In Section~\ref{sec:analysio} we study the numerical impact of
second-order power corrections on various $\bar B\to X_u\,l^-\bar\nu$
partial rates and find their effects to be tiny. It is therefore a safe 
approximation to neglect hadronic power corrections of order $1/m_b^3$ or
higher. The only possible exception to this conclusion relates to the 
so-called weak annihilation terms in $\bar B\to X_u\,l^-\bar\nu$ decay, 
which will be included in our error analysis.

\section{Inclusive semileptonic decays}
\label{sec:BuLnu}

All hadronic physics in $\bar B\to X_u\,l^-\bar\nu$ decays is encoded in the
hadronic tensor $W^{\mu \nu}$, which is defined via the discontinuity
of the forward $B$-meson matrix element of a correlator of two
flavor-changing weak currents $J^\mu=\bar u\,\gamma^\mu(1-\gamma_5)\,b$. 
Explicitly, 
\begin{equation}
   W^{\mu\nu} = \frac{1}{2 M_B}\,\frac1\pi \,\mbox{Im}\,\langle \bar B(v)| 
   i\!\int d^4x\, e^{iq\cdot x} \,\mbox{T}
   \left\{ J^{\dagger \mu}(0),J^\nu(x) \right\}
   |\bar B(v) \rangle \,,
\end{equation}
where $v$ is the $B$-meson velocity and $q$ the momentum carried by
the lepton pair. The hadronic tensor can be decomposed into five structure
functions $W_i$, which are the coefficients of the five possible
Lorentz structures built out of two independent 4-vectors. Typical choices
for these two vectors are $q$ and $v$, $p$ and $v$, etc. Here, as above, 
$p=m_b v-q$ is the momentum of the jet of light particles into which the
$b$ quark decays. In principle, all choices are equivalent, and it is
solely a matter of convenience which basis one picks. 

The triple differential decay rate can then be expressed in terms of
kinematical prefactors and the functions $W_i$. It is a known fact that
the total decay rate is proportional to five powers of the $b$-quark
mass. Further sensitivity to $m_b$ is picked up for partial decay
rates by the kinematical cuts. For example, cutting on the leptonic
invariant mass $q^2 > q_0^2$ introduces roughly five additional powers,
and the resulting partial decay rate is proportional to $(m_b)^a$ with
$a\approx 10$ \cite{Neubert:2000ch,Neubert:2001ib}. This is the reason why
theoretical predictions were typically made for event fractions, so
that at least the five powers of $m_b$ in the total rate drop out. For
practical purposes, however, this procedure presents no advantage as
the value of the total decay rate cannot be measured. Furthermore, the
$m_b$ dependence of the total rate is clearly related to the $m_b$
dependence of partial rates, and it is important to take
this correlation into account when combining calculations of event fractions
with those of the total decay rate. In
Section~\ref{sec:analysio}, where we present theoretical predictions,
we will thus focus directly on predictions for 
partial decay rates, not event fractions. Note that 
information about $m_b$ enters the triple differential decay
rate in two ways, via the hadronic structure
functions $W_i$ and through their kinematical prefactors. Whether or not 
$m_b$ appears explicitly in the prefactors depends on the decomposition of
$W^{\mu \nu}$, i.e., on the choice of vectors used to form the five
possible Lorentz structures.

A very useful set of 4-vectors turns out to be $(v,n)$, where $n$ is a
light-like vector in the direction of the jet of light
particles. In SCET, $n$ denotes the direction of
the collinear particles in the jet, which is typically set to be along the
$z$-axis. The normalization is chosen such that $v\cdot n = 1$, so
that $n^\mu=(1,0,0,1)$ in the rest frame of the $B$ meson. The conjugate
direction to $n$ is denoted by $\bar n^\mu=(1,0,0,-1)$ and marks the
direction of the photon in $\bar B\to X_s\gamma$ decay, or the
direction of the lepton pair in 
$\bar B\to X_u\,l^-\bar\nu$ decay. We then decompose
\begin{eqnarray}
   W^{\mu \nu}
   &=& (n^\mu v^\nu + n^\nu v^\mu -g^{\mu \nu}
    - i \epsilon^{\mu \nu \alpha \beta} n_\alpha v_\beta )\,\tilde W_1
    - g^{\mu \nu}\,\tilde W_2 \nonumber\\
   &&\mbox{}+ v^\mu v^\nu\,\tilde W_3
    + (n^\mu v^\nu + n^\nu v^\mu)\,\tilde W_4
    + n^\mu n^\nu\,\tilde W_5 \,.
\end{eqnarray}
The structure functions $\tilde W_i$ all have mass dimension
$-1$ in this basis. In terms of the $\tilde W_i$ functions the triple
differential decay rate reads
\begin{eqnarray}\label{eq:tripleRate}
   \frac{d^3\Gamma_u}{dP_+\,dP_-\,dP_l}
   &=& \frac{G_F^2|V_{ub}|^2}{16\pi^3}\,U_y(\mu_h,\mu_i)\,(M_B-P_+)\,
    \Big[ (P_- -P_l)(M_B-P_- +P_l-P_+)\,\F_1 \nonumber\\
   &&\mbox{}+ (M_B-P_-)(P_- -P_+)\,\F_2 + (P_- -P_l)(P_l-P_+)\,\F_3 \Big] \,,
\end{eqnarray}
where we have collected the relevant combinations of $\tilde W_i$ into the
three functions
\begin{equation}
   U_y(\mu_h,\mu_i)\,\F_1 = \tilde W_1 \,, \qquad 
   U_y(\mu_h,\mu_i)\,\F_2 = \frac{\tilde W_2}{2} \,, \qquad 
   U_y(\mu_h,\mu_i)\,\F_3 = \left( \frac{y}{4}\,\tilde W_3 + \tilde W_4
    + \frac{1}{y}\,\tilde W_5 \right)
\end{equation}
and defined a new kinematical variable
\begin{equation}\label{ydef}
   y = \frac{P_- -P_+}{M_B-P_+} \,,
\end{equation}
which can take values $0\le y\le 1$.
The leading evolution factor $U_y(\mu_h,\mu_i)$ has been factored out in 
(\ref{eq:tripleRate}) for convenience, as we have done earlier in 
(\ref{BsG:general}). The function $U_y(\mu_h,\mu_i)$ differs from the 
corresponding function in $\bar B\to X_s\gamma$ decay by a $y$-dependent
factor, 
\begin{equation}\label{eq:SudakovFactor}
   U_y(\mu_h,\mu_i) = U(\mu_h,\mu_i)\,y^{-2a_\Gamma(\mu_h,\mu_i)} \,,
\end{equation}
where the function $a_\Gamma$ in the exponent is related to the cusp
anomalous dimension and is given in Appendix~\ref{apx:Sudakovs}.

Eq.~(\ref{eq:tripleRate}) for the triple differential rate is exact.
Note that there is
no reference to the $b$-quark mass at this point. The only dependence
on $m_b$ is through the structure functions $\F_i(P_+,y)$ (via hard matching
corrections and via the moment constraints on the shape function $\hat S$),
which are independent of the leptonic variable
$P_l$. The fact that the total decay rate $\Gamma_u$ is proportional
to $m_b^5$ is not in contradiction with (\ref{eq:tripleRate}). It is 
instructive to demonstrate how these five powers of $m_b$ are recovered in 
our approach. At tree level and leading
power the functions $\F_2$ and $\F_3$ vanish, while $\F_1=\hat S(P_+)$. 
Integrating over the full range of $P_l$ and $P_-$ builds up
five powers of $(M_B-P_+)$. For the purpose of illustration, let us
rename the $P_+$ variable to $\hat\omega$ in the last integration, so that 
the total decay rate is given as
\begin{eqnarray}\label{ex:totalRateOPErecovery}
   \Gamma_u 
   &=& \frac{G_F^2|V_{ub}|^2}{192\pi^3} \int_0^{M_B}\!d\hat\omega\,
    (M_B-\hat\omega)^5\,\hat S(\hat\omega)
   = \frac{G_F^2|V_{ub}|^2}{192\pi^3} \int_{-m_b}^{M_B-m_b}\!\! 
    d\omega\,(m_b+\omega)^5\,S(\omega) \nonumber\\
   &=& \frac{G_F^2|V_{ub}|^2}{192\pi^3} (m_b+\langle\omega\rangle)^5
    \left[ 1 + O\left(\frac{1}{m_b^2}\right) \right] .
\end{eqnarray}
At tree level, the first moment of the shape function $S(\omega)$ vanishes.
Beyond tree level this is no longer the
case, and the average $\langle\omega\rangle$ depends on the size of
the integration domain. The above observation motivates the use of the
shape-function scheme \cite{Bosch:2004th}, in which the $b$-quark mass is 
defined as $m_b^{\rm SF}=m_b^{\rm pole}+\langle\omega\rangle+O(1/m_b)$.  
After this is done, (\ref{ex:totalRateOPErecovery})
recovers the form of the conventional OPE result.

\begin{figure}[t]
\begin{center}
\epsfig{file=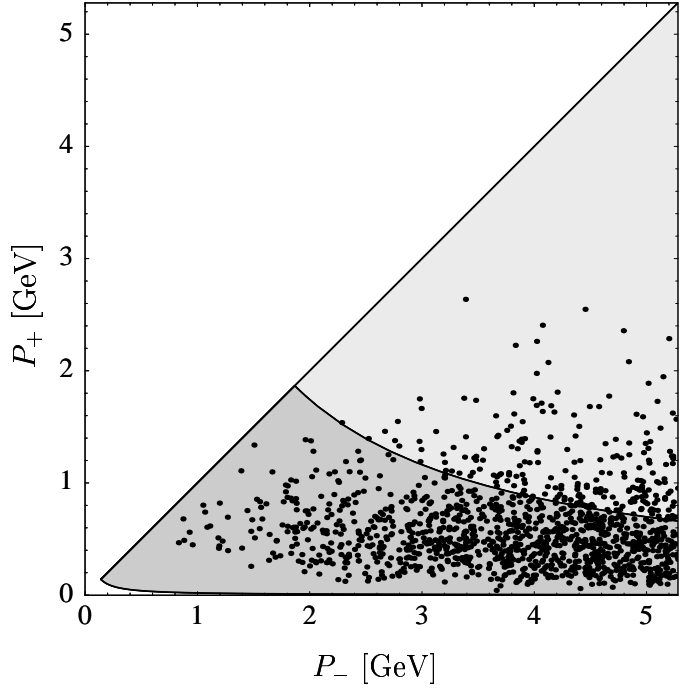, width=9cm}
\caption{\label{fig:scatterPlot}
The hadronic phase space in $P_+$ and $P_-$. The light gray region contains 
background from $\bar B\to X_c\,l^-\bar\nu$ decays, while the dark gray region 
is only populated by $\bar B\to X_u\,l^-\bar\nu$ events. The line separating 
the two regions is the contour where $M_X^2=P_+ P_-=M_D^2$.
Each point represents a $\bar B\to X_u\,l^-\bar\nu$ event 
in a Monte-Carlo simulation using the results of this paper. While the 
shape-function region of large $P_-$ and small $P_+$ is highly populated, 
there is not a single event with $P_+$ larger than 3\,GeV out of the 1300 
events generated.}
\end{center}
\end{figure}

Eq.~(\ref{eq:tripleRate}) and the above argument tell us that the
differential rate is a priori rather insensitive to the $b$-quark mass
in the endpoint region, where $P_+$ (and therefore $\langle\hat\omega\rangle$) 
is a small quantity. Only when the rates are integrated over a sufficiently
wide domain, so that shape-function integrals can be approximated using
a moment expansion, a dependence on $m_b$ enters indirectly via the first
moment of the leading-order shape function. Likewise, a dependence on other
HQET parameters such as $\lambda_1$ enters via the sensitivity to higher
moments.

In the remainder of this section we present the various
contributions to the structure functions $\F_i$, following the same line 
of presentation as we did in the case of $\bar B\to X_s\gamma$ decay in 
Section~\ref{sec:BsGamma}. As before, while the resulting expressions 
are ``optimized'' for the shape-function region, they can
be used over the entire phase space and give the correct result for
the total decay rate up to corrections of $O(\alpha_s^2)$. In the
shape-function region, where $P_+$ is a small quantity, one may
organize each $\F_i$ as a series in inverse powers of $1/(M_B-P_+)$. No
assumption about the variable $y$ is made, which is treated as an
$O(1)$ quantity.\footnote{In the shape-function region, where $P_+\ll P_-$, we 
have $y\approx p_-/m_b$, which is the variable used in the leading-power
analysis in \cite{Bosch:2004th}.}
A preview of the results of our calculation is depicted in
Figure~\ref{fig:scatterPlot}, which shows 
an illustration of our prediction for the distribution of events in 
the plane $(P_+,P_-)$. 

\subsection{Leading-power factorization formula}
\label{sec:BuWleadPow}

The leading-power expressions for the hadronic structure functions
$W_i$ have been calculated in \cite{Bosch:2004th} at one-loop order in
renormalization-group improved perturbation theory. 
At this level $\F_2$ does not obtain a contribution,
whereas $\F_1$ and $\F_3$ do. Symbolically, they take the factorized form
$H_{ui}\,J\otimes\hat S$, consisting of hard functions $H_{ui}$ and the 
convolution of the jet function $J$ with the leading shape function $\hat S$. 
More precisely,
\begin{equation}\label{eq:BuLP}
   \F_i^{\rm (0)}(P_+,y)
   = H_{ui}(y,\mu_h) \int_0^{P_+}\!d\hat\omega\,y m_b\,
    J(y m_b(P_+-\hat\omega),\mu_i)\,\hat S(\hat\omega,\mu_i) \,,
\end{equation}
where the hard functions are given by
\begin{eqnarray}\label{eq:LeadHardFunction}
   H_{u1}(y,\mu_h)
   &=& 1 +\! \frac{C_F\alpha_s(\mu_h)}{4\pi} \!\left[\!
    - 4\ln^2\!\frac{y m_b}{\mu_h} + 10\ln\frac{y m_b}{\mu_h} - 4\ln y
    - \frac{2\ln y}{1-y} - 4L_2(1\!-\!y) - \frac{\pi^2}{6} -\! 12 \right]\! ,
    \nonumber\\   
   H_{u3}(y,\mu_h)
   &=& \frac{C_F\alpha_s(\mu_h)}{4\pi}\,\frac{2\ln y}{1-y} \,,
\end{eqnarray}
and $H_{u2}=0$.
As before, the differential decay rate is independent of the matching scales 
$\mu_h\sim m_b$ and $\mu_i\sim\sqrt{m_b\Lambda_{\rm QCD}}$.
The jet function $J$ has already been given in (\ref{BsG:jetfunc}). 
Note that the $b$-quark mass appears
only as the argument of logarithms, where it plays the role of setting 
the renormalization scale.

\subsection{Kinematical power corrections}

As in the case of $\bar B\to X_s\gamma$ decay, there is a class of power
corrections to the $\bar B\to X_u\,l^-\bar\nu$ decay distributions which
are small only because of the restriction to certain regions in phase space,
but which are not associated with new hadronic parameters. In the present 
case, these terms can be extracted from the one-loop expressions derived 
in \cite{DeFazio:1999sv}. They are then convoluted with the leading shape 
function. As previously, the scale separation that can be achieved for these 
power-suppressed terms is only approximate, and we thus assign a coupling
$\alpha_s(\bar\mu)$ with them, where the scale $\bar\mu$ is expected to be 
of order $\mu_i\sim\sqrt{m_b\Lambda_{\rm QCD}}$.

The resulting expressions for the structure functions can be written in
a compact form in terms of the variables $x$ and $y$ defined in 
(\ref{xdef}) and (\ref{ydef}). We find
\begin{eqnarray}\label{eq:fullDFN}
   \F_1^{\rm kin}(P_+,y)
   &=& \frac{1}{M_B-P_+}\,\frac{C_F\alpha_s(\bar\mu)}{4\pi}
    \int_0^{P_+}\!d\hat\omega\,\hat S(\hat\omega,\mu_i) \nonumber\\
   &&\times \left[ \frac{f_1(x,y)}{(1+x)^2 y(x+y)}
    - \frac{2 g_1(x,y)}{x(1+x)^2 y^2(x+y)} \ln\left(1+\frac{y}{x} \right)
    - \frac{4}{x} \ln\left( y+\frac{y}{x} \right) \right] , \nonumber\\
   \F_2^{\rm kin}(P_+,y)
   &=& \frac{1}{M_B-P_+}\,\frac{C_F\alpha_s(\bar\mu)}{4\pi}
    \int_0^{P_+}\!d\hat\omega\,\hat S(\hat\omega,\mu_i) \nonumber\\
   &&\times \left[ \frac{f_2(x,y)}{(1+x)^2y^2(x+y)}
    - \frac{2x\,g_2(x,y)}{(1+x)^2 y^3(x+y)} 
    \ln\left(1+\frac{y}{x}\right) \right] , \nonumber \\
   \F_3^{\rm kin}(P_+,y)
   &=& \frac{1}{M_B-P_+}\,\frac{C_F\alpha_s(\bar\mu)}{4\pi}
    \int_0^{P_+}\!d\hat\omega\,\hat S(\hat\omega,\mu_i) \nonumber \\
   &&\times \left[ \frac{f_3(x,y)}{(1+x)^2y^3(x+y)}
    + \frac{2 g_3(x,y)}{(1+x)^2y^4(x+y)} 
    \ln\left( 1+\frac{y}{x} \right) \right] \,,
\end{eqnarray}
where the functions $f_i$, $g_i$ are given by
\begin{eqnarray}
   f_1(x,y) &=& -9 y + 10 y^2 + x (-16+12y+6y^2) + x^2(13 y-12) \,,
    \nonumber\\
   g_1(x,y) &=& -2y^3 -2 x y^2(4+y) -x^2 y(12+4y+y^2) - 4x^3 (y+2)
    + 3 x^4(y-2) \,, \nonumber\\
   f_2(x,y) &=& y^2+xy(8+4y+y^2) +3x^2y(10+y)+x^3(12+19y)+10x^4 \,,
    \nonumber\\
   g_2(x,y) &=&  2y^2+4xy(1+2y)+x^2y(18+5y) + 6x^3(1+2y)+5x^4 \,, \nonumber\\
   f_3(x,y) &=& 2y^3(2y-11) +xy^2(-94+29y+2y^2) + 2x^2y(-72+18y+13y^2)
    \nonumber\\
   &&\mbox{}+ x^3(-72-42y+70y^2-3y^3) - 10x^4(6-6y+y^2) \,,\nonumber \\
   g_3(x,y) &=& 4y^4 -6x(y-5)y^3 - 4x^2y^2(-20+6y+y^2) 
    + x^3y(90-10y-28y^2+y^3) \nonumber\\
   &&\mbox{}+ x^4(36+36y-50y^2+4y^3) +5x^5(6-6y+y^2) \,.
\end{eqnarray}
The above formulae are the exact $O(\alpha_s)$ corrections to the
leading-power expression. This means that, when integrated over the entire 
phase space, they will give rise to the correct result for the total rate up 
to that order. In the shape-function region (where $P_+\ll P_-$)
the integrands in (\ref{eq:fullDFN}) can be expanded in powers of $1/m_b$ by 
counting $y=O(1)$ and $x=O(1/m_b)$.
Note that this organizes the $1/m_b$ expansion as an expansion
in powers of the hadronic variable $1/(M_B-P_+)$. The leading terms read
\begin{eqnarray}\label{eq:kinNLO}
   \F_1^{\rm kin(1)}(P_+,y)
   &=& \frac{1}{M_B-P_+}\,\frac{C_F\alpha_s(\bar\mu)}{4\pi}
    \int_0^{P_+}\!d\hat\omega\,\hat S(\hat\omega,\mu_i)
    \left[ 6 - \frac5y + \left( \frac{12}{y}-4 \right) \ln\frac{y}{x} \right]
    , \nonumber\\
   \F_2^{\rm kin(1)}(P_+,y)
   &=& \frac{1}{M_B-P_+}\,\frac{C_F\alpha_s(\bar\mu)}{4\pi}
    \int_0^{P_+}\!d\hat\omega\,\hat S(\hat\omega,\mu_i)
    \left[ \frac1y \right] , \nonumber\\
   \F_3^{\rm kin(1)}(P_+,y)
   &=& \frac{1}{M_B-P_+}\,\frac{C_F\alpha_s(\bar\mu)}{4\pi}
    \int_0^{P_+}\!d\hat\omega\,\hat S(\hat\omega,\mu_i) 
    \left[ 4 - \frac{22}{y} + \frac8y \ln\frac{y}{x} \right] .
\end{eqnarray}
Further accuracy can be achieved by adding the next-order corrections, 
for which we obtain
\begin{eqnarray}\label{eq:kinNNLO}
   \F_1^{\rm kin(2)}(P_+,y)
   &=& \frac{1}{(M_B-P_+)^2}\,\frac{C_F\alpha_s(\bar\mu)}{4\pi}
    \int_0^{P_+}\!d\hat\omega\,(P_+ -\hat\omega)\,\hat S(\hat\omega,\mu_i)
    \nonumber\\
   &&\times \left[ -12 + \frac{16}{y} + \frac{3}{y^2} 
    + \left( \frac{12}{y^2} - \frac{20}{y} + 6 \right) 
    \ln\frac{y}{x} \right] , \nonumber\\
   \F_2^{\rm kin(2)}(P_+,y)
   &=& \frac{1}{(M_B-P_+)^2}\,\frac{C_F\alpha_s(\bar\mu)}{4\pi}
    \int_0^{P_+}\!d\hat\omega\,(P_+ -\hat\omega)\,\hat S(\hat\omega,\mu_i)
    \nonumber\\
   &&\times \left[ 1 + \frac2y + \frac{7}{y^2} - \frac{4}{y^2} \ln\frac{y}{x}
    \right] , \nonumber\\
   \F_3^{\rm kin(2)}(P_+,y)
   &=& \frac{1}{(M_B-P_+)^2}\,\frac{C_F\alpha_s(\bar\mu)}{4\pi}
    \int_0^{P_+}\!d\hat\omega\,(P_+ -\hat\omega)\,\hat S(\hat\omega,\mu_i)
    \nonumber \\
   &&\times \left[ - 6 + \frac{69}{y} - \frac{64}{y^2}
    + \left( \frac{52}{y^2} - \frac{28}{y} \right) \ln\frac{y}{x} \right] .
\end{eqnarray}
In the various phase-space regions of interest to the determination of 
$|V_{ub}|$, the above terms (\ref{eq:kinNLO}) and (\ref{eq:kinNNLO}) 
approximate the full result (\ref{eq:fullDFN}) very well (see 
Section~\ref{sec:analysio} below).  

Let us comment here on a technical point already mentioned in the 
Introduction. When combined with the phase-space factors
in (\ref{eq:tripleRate}), the exact expressions for $\F_i^{\rm kin}$ in 
(\ref{eq:fullDFN}) are regular in the limit $P_-\to P_+$, corresponding to 
$y\to 0$. However, this feature is not automatically ensured when the
structure functions, but not the phase-space factors, are expanded about the 
heavy-quark limit. With our choice of the variables $x$ and $y$, we encounter
terms as singular as $1/y^n$ at $n$-th order in the expansion, as is obvious 
from the explicit expressions above. Phase space
scales like $y^2$ in the limit $y\to 0$ (note that $P_l\to P_+$ as 
$P_-\to P_+$ because of (\ref{eq:simplePhaseSpace})), so that the results 
(\ref{eq:kinNLO}) and (\ref{eq:kinNNLO}) can be applied without encountering
any kinematical singularities. In order to achieve this, it was crucial to 
define the variable
$y$ in the way we did in (\ref{ydef}). We emphasize this point because
straightforward application of the technology of SCET and HQET developed 
in \cite{Bosch:2004cb,Lee:2004ja,Beneke:2004in} would give an expansion of 
the structure functions $\F_i$ in powers of $1/p_-$, whereas phase space
is proportional to $4\vec{p}^{\,2}=(p_- -p_+)^2\propto y^2$. In the 
kinematical region where $p_+<0$, which is allowed due to off-shell effects in
the $B$ meson, this leads to singularities as $p_-\to 0$. In order to
avoid these singularities, we have reorganized the SCET expansion as an 
expansion in $1/(p_- -p_+)$ instead of $1/p_-$, where $|p_+|\ll p_-$ in the 
shape-function region.

\subsection{Subleading shape-function contributions}
\label{sec:BuWsubSF}

The contributions from subleading shape functions to arbitrary 
$\bar B\to X_u\,l^-\bar\nu$ decay distributions have been derived (at tree 
level) in \cite{Bosch:2004cb,Lee:2004ja,Beneke:2004in}. 
The results involve the same set of subleading shape functions as previously 
discussed in Section~\ref{subsec:BsG:SSF}. 
Again, the structure function $\F_2$ does not obtain a contribution, while
\begin{eqnarray}
   \F_1^{\rm hadr(1)}(P_+,y)
   &=& \frac{1}{M_B-P_+}
    \left[ (\bar\Lambda-P_+)\,\hat S(P_+) + \hat t(P_+) 
    + \frac{\hat u(P_+) - \hat v(P_+)}{y} \right] , \nonumber\\
   \F_3^{\rm hadr(1)}(P_+,y)
   &=& \frac{1}{M_B-P_+}\,\frac{2}{y}
    \left[ - (\bar\Lambda-P_+)\,\hat S(P_+) - 2\hat t(P_+) 
    + \frac{\hat t(P_+) + \hat v(P_+)}{y} \right] .
\end{eqnarray}
At this point we recall the discussion of Section~\ref{subsec:BsG:SSF}, where 
we have argued that the $\bar B\to X_s\gamma$ photon spectrum should be used 
to fit the function $\hat\S$ of (\ref{def:ScriptS}), which is defined to
be a linear combination of the leading shape function $\hat S$ and the
subleading shape functions $\hat t$, $\hat u$, $\hat v$. When the above 
results are rewritten in terms of the new function $\hat\S$
nothing changes in the expressions for $\F_i^{\rm (0)}$ except for
the simple replacement $\hat S\to \hat\S$, which we from now on
assume. At the level of subleading shape functions 
$\F_2^{\rm hadr(1)}=0$ and $\F_3^{\rm hadr(1)}$ remain unchanged, while
\begin{equation}\label{upsy}
   \F_1^{\rm hadr(1)}(P_+,y) = \frac{1}{M_B-P_+}
   \left[ - (\bar\Lambda-P_+)\,\hat\S(P_+) + 2\hat t(P_+) 
   + \left[ \hat u(P_+) - \hat v(P_+) \right]
   \left( \frac{1}{y} - 1 \right) \right] .
\end{equation}
It follows that there reside some linear combinations of subleading shape
functions in the triple differential decay rate that cannot be
extracted from information on the photon spectrum in $\bar B\to X_s\gamma$
decays. In the end, this dependence gives rise to a theoretical
uncertainty.

\subsection{Residual hadronic power corrections}
\label{sec:MWterms}

In analogy with our treatment for the case of $\bar B\to X_s\gamma$ decay, we
start from the expressions for the $1/m_b^2$ corrections to the triple
differential $\bar B\to X_u\,l^-\bar\nu$ decay rate obtained by applying the
OPE to the hadronic tensor \cite{Blok:1993va,Manohar:1993qn}.
Converting these results into the $(v,n)$ basis and
changing variables from $v\cdot q$ and $q^2$ to $p_+=n\cdot p$ and
$p_-=\bar n\cdot p$, we find
\begin{eqnarray}\label{eq:OPEres}
   \tilde W_1^{(2)}
   &=& \delta(p_+) \left( 1 + \frac{2\lambda_1-3\lambda_2}{3p_-^2} \right) 
    + \delta'(p_+) \left( \frac{2\lambda_1-3\lambda_2}{3p_-} 
    - \frac{5\lambda_1+15\lambda_2}{6m_b} \right) 
    - \delta''(p_+)\,\frac{\lambda_1}{6} \,, \nonumber \\ 
   \tilde W_2^{(2)}
   &=& \delta(p_+) \left( - \frac{4\lambda_1-6\lambda_2}{3p_-^2} \right) , \\
   \frac{y}{4}\,\tilde W_3^{(2)} &+& \tilde W_4^{(2)}
    + \frac{1}{y}\,\tilde W_5^{(2)}
   = \frac{\delta(p_+)}{p_-} \left( \frac{2\lambda_1+12\lambda_2}{3p_-}
    - \frac{4\lambda_1+9\lambda_2}{3m_b} \right)
    + \frac{\delta'(p_+)}{p_-} \left( \frac{2\lambda_1}{3} 
    + 4\lambda_2 \right) . \nonumber
\end{eqnarray}
The desired $1/(M_B-P_+)^2$ corrections to the structure functions
$\F_i$ can then be extracted by expanding the leading and subleading
contributions $\F_i^{\rm (0)}$ and $\F_i^{\rm hadr(1)}$ in terms of their
moments in (\ref{SSF:moments}), and by subtracting the results from
(\ref{eq:OPEres}). Following the same procedure as in 
Section~\ref{subsec:BsGHQE} to express the remaining power corrections in 
terms of the leading shape function, we obtain
\begin{eqnarray}
   \F_1^{\rm hadr(2)}(P_+,y)
   &=& \frac{1}{(M_B-P_+)^2} \left( \frac{4\lambda_1-6\lambda_2}{3y^2} 
    - \frac{\lambda_1+3\lambda_2}{3} \right) \hat S(P_+) \,,
    \nonumber\\
   \F_2^{\rm hadr(2)}(P_+,y)
   &=& \frac{1}{(M_B-P_+)^2}
    \left( \frac{-2\lambda_1+3\lambda_2}{3y^2} \right) \hat S(P_+) \,,
    \nonumber\\
   \F_3^{\rm hadr(2)}(P_+,y)
   &=& \frac{1}{(M_B-P_+)^2} \left( \frac{4\lambda_1+24\lambda_2}{3y^2} 
    - \frac{4\lambda_1+9\lambda_2}{3y} \right) \hat S(P_+) \,.
\end{eqnarray}
These expressions remain unchanged when the shape function $\hat\S$ is used
instead of $\hat S$.

\subsection{Weak annihilation contributions}
\label{Sec:WA}

In the OPE calculation several contributions appear at third order in the
power expansion: $1/m_b$ corrections to the kinetic and chromo-magnetic
operators, the Darwin and spin-orbit terms, and weak annihilation 
contributions. The Darwin and spin-orbit terms correspond to the forward 
$B$-meson matrix elements of (light) flavor-singlet 
operators \cite{Uraltsev:1999rr}. The corresponding HQET parameters 
$\rho^3_D$ and $\rho_{LS}^3$ can in principle be extracted from moments of 
inclusive $\bar B\to X_c\,l^-\bar\nu$ decay spectra. They are insensitive 
to the flavor of the spectator quark inside the $B$ meson. The weak 
annihilation contribution, on the other hand, results from four-quark 
operators with flavor non-singlet structure.
Graphically, this contribution corresponds to a process in which
the $b$ and $\bar u$ quark annihilate into a $W^-$. Weak annihilation terms 
come with a phase-space enhancement factor of $16\pi^2$ and so are potentially 
more important than other power corrections of order $1/m_b^3$.
Because of the
flavor dependence, these contributions can effect neutral and charged
$B$ mesons differently \cite{Bigi:1993bh}. 
One choice of basis for the
corresponding four-quark operators is \cite{Neubert:1996we}
\begin{equation}
   \langle\bar B|\bar b_L\gamma_\mu u_L\,\bar u_L\gamma^\mu b_L|\bar B\rangle
    = \frac{f_B^2 M_B^2}{4}\,B_1 \,, \qquad
   \langle\bar B|\bar b_R u_L\,\bar u_L b_R|\bar B\rangle
    = \frac{f_B^2 M_B^2}{4}\,B_2 \,,
\end{equation}
where $f_B$ is the $B$-meson decay constant, and $B_i$ are hadronic
parameters. In the vacuum saturation approximation they are given by
$B_1=B_2=1$ for charged $B$ mesons and $B_1=B_2=0$ for neutral ones. 
The total semileptonic rate is proportional to the
difference $(B_2-B_1)$, which implies that the weak annihilation contribution 
would vanish in this approximation. Currently, only rough estimates are 
available for the magnitude of the deviation of this difference from zero. 
The resulting effect on the total branching ratio is 
\cite{Voloshin:2001xi}
\begin{equation}
   \delta B(\bar B\to X_u\,l^-\bar\nu)\approx
   3.9 \left( \frac{f_B}{0.2\,{\rm GeV}} \right)^2
   \left( \frac{B_2-B_1}{0.1} \right) |V_{ub}|^2 \,.
\end{equation}
Again, we expect this effect to be different for charged and neutral $B$
mesons. The most important feature of weak annihilation is that it is 
formally concentrated
at the kinematical point where all the momentum of the heavy quark is
transferred to the lepton pair \cite{Bigi:1993bh}. At the parton level
this implies that the corresponding contribution is proportional to
$\delta(q^2-m_b^2)$. It is therefore included in
every cut that includes the $q^2$ endpoint, and its effect is
independent of the specific form of the cut. 

We suggest two different strategies to control this effect. The
first is to include it in the error estimate as a constant contribution 
proportional to the total rate. A recent study \cite{TomsThesis} puts
a limit on this effect
of $\pm 1.8\%$ on the total rate (at 68\% confidence level) by analyzing 
CLEO data. The second one is to impose a cut $q^2\le q_{\rm max}^2$, 
thus avoiding the region where the weak annihilation contribution is
concentrated. The maximal value of $q^2$ is $(M_B-M_\pi)^2$, but one
must exclude a larger region of phase space, such that the {\em excluded\/} 
contribution to the decay rate at large $q^2$ 
(corresponding to a region near the origin in the 
$(P_-,P_+)$ plane) can be reliably calculated. In our
numerical analysis, we will study the effect of a cut $q^2\le(M_B-M_D)^2$,
which satisfies this criterion.

For completeness, we note that even after the weak annihilation contribution
near maximum $q^2$ has been removed, there could in principle exist other, 
flavor-specific contributions to the semileptonic decay amplitudes that are
different for charged and neutral $B$ mesons. The leading terms of this kind  
contribute at order $1/m_b$ in the shape-function region and are parameterized 
by a set of four-quark subleading shape functions 
\cite{Bosch:2004cb,Lee:2004ja,Beneke:2004in}. Model estimates of these 
contributions show 
that they are very small for all observables considered for an extraction of 
$|V_{ub}|$ \cite{Beneke:2004in,Neubert:2004cu}. If only flavor-averaged decay
rates are measured, the effects of four-quark subleading shape functions can 
be absorbed entirely by a redefinition of the functions $\hat u(\hat\omega)$ 
and $\hat v(\hat\omega)$ \cite{Bosch:2004cb}, without affecting the moment 
relations in (\ref{SSF:moments}).

\section{Shape-function parameterizations}
\label{sec:strucfunc}

Hadronic-physics effects enter the description of inclusive decay
rates via non-perturbative shape functions. Perturbation theory cannot
tell us much about the local form of these functions, but moments of
them are calculable provided that the domain of integration is much
larger than $\Lambda_{\rm QCD}$. Since the shape functions contain
information about the internal structure of the 
$B$ meson, knowledge of them relates directly to the determination
of the $b$-quark mass $m_b$, the kinetic-energy parameter $\lambda_1$, and 
in principle the matrix elements of higher-dimensional
operators. Improved measurements of the shape of the $\bar B\to X_s\gamma$
photon spectrum will therefore lead directly to a more precise
determination of HQET parameters. This argument can be 
turned around to constrain the leading shape function using
knowledge of $m_b$ and $\lambda_1$ from other physical processes such
as a $b\to c$ moment analysis \cite{Aubert:2004aw}. We emphasize,
however, that there are obviously infinitely many locally different
functions that have identical first few moments. In this section we
present a few functional forms that can be used to model the shape
functions and to fit the current experimental data.

To achieve stringent constraints on the leading shape function 
a precise definition of the HQET
parameters is required. It is a well-known fact that the pole-mass
scheme introduces uncontrollable ambiguities. To avoid these
uncertainties several short-distance definitions have been proposed,
such as the $\overline{\rm MS}$ scheme, the potential-subtraction
scheme \cite{Beneke:1998rk}, the $\Upsilon(1S)$ scheme
\cite{Hoang:1998hm}, the kinetic scheme \cite{Bigi:1996si}, or the
shape-function scheme \cite{Bosch:2004th}. While the decay rates are of 
course independent of the particular choice, it is advantageous to use a 
mass scheme that is designed for the physics problem at hand. 
In the case of inclusive
$B$ decays into light particles, this is the shape-function scheme.

\subsection{Models for the leading shape function}

Model-independent constraints on the shape function $\hat S(\hat\omega,\mu_i)$ 
can be derived by analyzing moments defined with an upper limit of 
integration $\hat\omega_0$, i.e.\
\begin{equation}\label{Def:Moments}
   M_N(\hat\omega_0,\mu_i)\equiv \int_0^{\hat\omega_0}\!d\hat\omega\,
   \hat\omega^N\,\hat S(\hat\omega,\mu_i) \,.
\end{equation}
For practical applications,
$\hat\omega_0$ should be taken of order the size of the window where
the $\bar B\to X_s\gamma$ photon spectrum is experimentally accessible,
$\hat\omega_0=M_B-2E_\gamma^{\rm min}$ with 
$E_\gamma^{\rm min}\approx 1.8$\,GeV. These moments can be expanded in
terms of matrix elements of local operators as long as $\hat\omega_0$
is large compared to $\Lambda_{\rm QCD}$. In the 
shape-function scheme, HQET parameters are defined to all orders in 
perturbation theory through ratios of such moments, e.g.\ 
\cite{Bosch:2004th}
\begin{eqnarray}\label{MomentRelations}
   \frac{M_1(\mu_f+\bar\Lambda(\mu_f,\mu_i),\mu_i)}
        {M_0(\mu_f+\bar\Lambda(\mu_f,\mu_i),\mu_i)}
   &=& \bar\Lambda(\mu_f,\mu_i) \,, \nonumber\\
   \frac{M_2(\mu_f+\bar\Lambda(\mu_f,\mu_i),\mu_i)}
        {M_0(\mu_f+\bar\Lambda(\mu_f,\mu_i),\mu_i)}
   &=& \frac{\mu_\pi^2(\mu_f,\mu_i)}{3} + \bar\Lambda^2(\mu_f,\mu_i) \,. 
\end{eqnarray}
Here, the factorization scale $\mu_f\gg\Lambda_{\rm QCD}$ is related to 
the size of the integration domain via the implicit equation 
$\hat\omega_0=\mu_f+\bar\Lambda(\mu_f,\mu_i)$. In practice $\mu_f$ is close to 
the intermediate scale $\mu_i$. At tree level, the relations between
parameters in the shape-function scheme and the pole scheme are
$\bar\Lambda(\mu_f,\mu_i)=\bar\Lambda^{\rm pole}$ and
$\mu_\pi^2(\mu_f,\mu_i)=-\lambda_1$. The corresponding relations at 
one- and two-loop order have been worked out in \cite{Bosch:2004th} and
\cite{Neubert:2004sp}, respectively. These relations allow us to obtain
precise determinations of $\bar\Lambda(\mu_f,\mu_i)$ and
$\mu_\pi^2(\mu_f,\mu_i)$ from other physical processes.

For reference purposes, it is helpful to quote values for $\bar\Lambda$
and $\mu_\pi^2$ using only a single scale $\mu_*$ instead of two
independent scales $\mu_f$ and $\mu_i$. To one-loop order, these parameters 
can be related to those determined from the moments via \cite{Bosch:2004th}
\begin{eqnarray}\label{eq:atMuStar}
   \bar\Lambda(\mu_*,\mu_*)
   &=& \bar\Lambda(\mu_f,\mu_i) 
    + \mu_* \frac{C_F\alpha_s(\mu_*)}{\pi} 
    - \mu_f \frac{C_F\alpha_s(\mu_i)}{\pi} \left[ 1
    - 2\,\bigg( 1 - \frac{\mu_\pi^2(\mu_f,\mu_i)}{3\mu_f^2} \bigg)
    \ln\frac{\mu_f}{\mu_i} \right] , \\
   \mu_\pi^2(\mu_*,\mu_*)
   &=& \mu_\pi^2(\mu_f,\mu_i) 
    \left[ 1+\frac{C_F\alpha_s(\mu_*)}{2\pi} 
    - \frac{C_F\alpha_s(\mu_i)}{\pi} \left( \frac12 + 3 
    \ln\frac{\mu_f}{\mu_i} \right) \right]
    + 3\mu_f^2\,\frac{C_F\alpha_s(\mu_i)}{\pi}\,\ln\frac{\mu_f}{\mu_i} \,,
    \nonumber
\end{eqnarray}
where we have neglected higher-dimensional operator matrix elements
that are suppressed by inverse powers of $\mu_f$. A typical choice for
the scale $\mu_*$ is 1.5\,GeV, which we will use as the reference scale 
throughout this work.
It will be convenient to connect the parameter $\bar\Lambda$
extracted from the first moment of the shape function with a low-scale
subtracted quark-mass definition referred to as the ``shape-function'' mass. 
Following \cite{Bosch:2004th}, we define
\begin{equation}
   m_b(\mu_f,\mu_i)\equiv M_B - \bar\Lambda(\mu_f,\mu_i) \,.
\end{equation}

The general procedure for modeling the leading shape function $\hat
S(\hat\omega,\mu_i)$ from a given functional form $F(\hat\omega)$ is
as follows. The shape of $F(\hat\omega)$ is assumed to be tunable so
that it can be used to fit the $\bar B\to X_s\gamma$ photon spectrum. Only
the norm of the shape function is fixed theoretically. Note that
the moment relations (\ref{MomentRelations}) are insensitive to the
norm, so that formulae for $\bar\Lambda$ and $\mu_\pi^2$ follow
directly from the functional form of $F(\hat\omega)$. Examples of such
formulae will be given below. We define moments
$M_N^{[F]}(\hat\omega_0)$ of $F$ in analogy with
(\ref{Def:Moments}). The first relation in (\ref{MomentRelations})
implies that for a given $\hat\omega_0$ the factorization scale is
\begin{equation}
   \mu_f = \hat\omega_0
   - \frac{M_1^{[F]}(\hat\omega_0)}{M_0^{[F]}(\hat\omega_0)} \,.
\end{equation}
Now that $\mu_f$ is known, the norm is determined by requiring that
the zeroth moment of the shape function is \cite{Bosch:2004th}
\begin{equation} \label{SFnorm}
   M_0(\hat\omega_0,\mu_i) = 1 - \frac{C_F \alpha_s(\mu_i)}{\pi} 
   \left( \ln^2 \frac{\mu_f}{\mu_i} + \ln \frac{\mu_f}{\mu_i} 
    + \frac{\pi^2}{24} \right)
   + \frac{C_F\alpha_s(\mu_i)}{\pi} \left( \ln\frac{\mu_f}{\mu_i} 
    - \frac12 \right) \frac{\mu_\pi^2(\mu_f,\mu_i)}{3\mu_f^2} + \dots \,.
\end{equation}
It follows that $[M_0(\hat\omega_0,\mu_i)/M_0^{[F]}(\hat\omega_0)]\,
F(\hat\omega)$ serves as a model of $\hat S(\hat\omega,\mu_i)$ or 
$\hat\S(\hat\omega,\mu_i)$. 

\begin{figure}[t]
\begin{center}
\epsfig{file=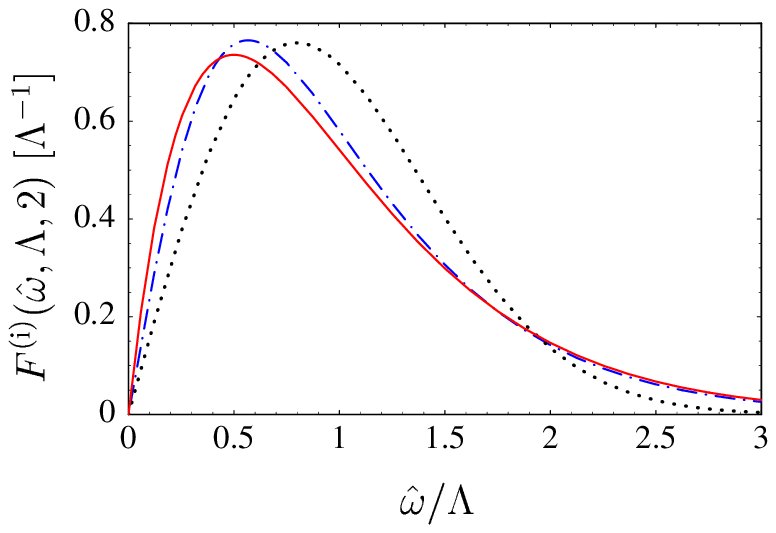, width=8cm}\hspace{3mm}%
\epsfig{file=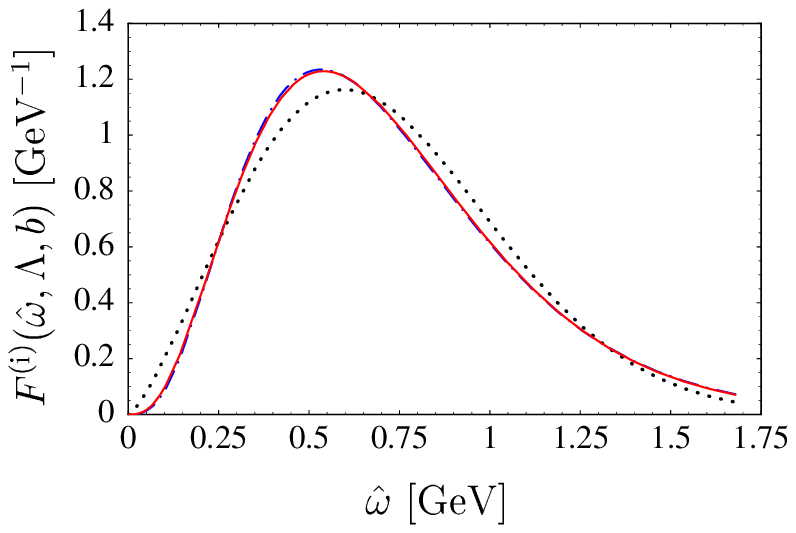, width=8cm}
\caption{\label{fig:SFmodels}
{\sc Left:} Different functional forms for the leading shape function. We 
show $F^{\rm (exp)}(\hat\omega,\Lambda,2)$ (solid), 
$F^{\rm (gauss)}(\hat\omega,\Lambda,2)$ (dotted), and 
$F^{\rm (hyp)}(\hat\omega,\Lambda,2)$ (dash-dotted) as functions of the 
ratio $\hat\omega/\Lambda$. 
{\sc Right:} The same functions with the parameters $\Lambda$ and $b$
tuned such that $m_b(\mu_*,\mu_*)=4.61$\,GeV and
$\mu_\pi^2(\mu_*,\mu_*) = 0.2$ GeV$^2$. See text for explanation.}
\end{center}
\end{figure}

We now suggest three two-parameter models for the leading-order
shape function based on an exponential-type function $F^{\rm (exp)}$, a
gaussian-type function $F^{\rm (gauss)}$, and hyperbolic-type function
$F^{\rm (hyp)}$.
We use two parameters that can be tuned to fit the photon spectrum: a
dimensionful quantity $\Lambda$ which coincides with the position of
the average $\langle \hat\omega \rangle$, and a positive number $b$
which governs the behavior for small $\hat\omega$. The functions we propose 
are
\begin{eqnarray}\label{SF:threeModels}
   F^{\rm (exp)}(\hat\omega;\Lambda,b)
   &=& \frac{N^{\rm (exp)}}{\Lambda}
    \left( \frac{\hat\omega}{\Lambda} \right)^{b-1}
    \exp\left( - d_{\rm (exp)} \frac{\hat\omega}{\Lambda} \right) , 
    \nonumber\\
   F^{\rm (gauss)}(\hat\omega;\Lambda,b)
   &=& \frac{N^{\rm (gauss)}}{\Lambda}
    \left( \frac{\hat\omega}{\Lambda} \right)^{b-1}
    \exp\left( - d_{\rm (gauss)} \frac{\hat\omega^2}{\Lambda^2} \right) , 
    \nonumber\\
   F^{\rm (hyp)}(\hat\omega;\Lambda,b)
   &=& \frac{N^{\rm (hyp)}}{\Lambda}
    \left( \frac{\hat\omega}{\Lambda} \right)^{b-1}
    \cosh^{-1}\left( d_{\rm (hyp)} \frac{\hat\omega}{\Lambda} \right) .
\end{eqnarray}
For convenience, we normalize these functions to unity. The parameters 
$d_{(i)}$ are determined by the choice $\Lambda=\langle\hat\omega\rangle$. 
We find
\begin{equation}
\begin{aligned}
   N^{\rm (exp)} &= \frac{d_{\rm (exp)}^b}{\Gamma(b)} \,, 
    & d_{\rm (exp)} &= b \,, \\
   N^{\rm (gauss)} &= \frac{2 d_{\rm (gauss)}^{b/2}}{\Gamma(b/2)} \,, 
    & d_{\rm (gauss)}
    &= \left(\frac{\Gamma(\frac{1+b}{2})}{\Gamma(\frac{b}{2})}\right)^2 \,,
    \\
   N^{\rm (hyp)} &= \frac{[4 d_{\rm (hyp)}]^b}{2\Gamma(b) 
    \left[ \zeta(b,\frac14) - \zeta(b,\frac34) \right]} \,, \quad
    & d_{\rm (hyp)}
    &= \frac{b}{4}\, 
    \frac{\zeta(1+b,\frac14) - \zeta(1+b,\frac34)}
         {\zeta(b,\frac14) - \zeta(b,\frac34)} \,,
\end{aligned}
\end{equation}
where $\zeta(b,a)=\sum_{k=0}^\infty (k+a)^{-b}$ is the generalized
Riemann zeta function.
An illustration of the different functional forms is given on the
left-hand side in Figure~\ref{fig:SFmodels}. We show a plot with the
choice $b=2$, corresponding to a linear onset for small $\hat\omega$. 

For the first two models, analytic expressions for the HQET parameters
$\bar\Lambda$ and $\mu_\pi^2$ are available. Following the discussion
above, we compute the moments on the interval $[0,\hat\omega_0]$ and
find for the exponential form $F^{\rm (exp)}(\hat\omega;\Lambda,b)$
\begin{eqnarray}\label{parsExp}
   \bar\Lambda(\mu_f,\mu_i)
   &=& \frac{\Lambda}{b}\,\frac{\Gamma(1+b)
             - \Gamma(1+b,\frac{b\,\hat\omega_0}{\Lambda})}
            {\Gamma(b) - \Gamma(b,\frac{b\,\hat\omega_0}{\Lambda})} \,,
    \nonumber\\
   \mu_\pi^2(\mu_f,\mu_i)
   &=& 3 \left[ \frac{\Lambda^2}{b^2}\, 
    \frac{\Gamma(2+b) - \Gamma(2+b,\frac{b\,\hat\omega_0}{\Lambda})}
         {\Gamma(b) - \Gamma(b, \frac{b\,\hat\omega_0}{\Lambda})}\,
    - \bar\Lambda(\mu_f,\mu_i)^2 \right] ,
\end{eqnarray}
where $\mu_f=\hat\omega_0-\bar\Lambda(\mu_f,\mu_i)$. A similar calculation for 
the gaussian form $F^{\rm (gauss)}(\hat\omega;\Lambda,b)$ yields
\begin{eqnarray}\label{parsGauss}
   \bar\Lambda(\mu_f,\mu_i)
   &=& \frac{\Lambda}{\sqrt{d_{\rm (gauss)}}}\, 
    \frac{\Gamma(\frac{1+b}{2})
     - \Gamma(\frac{1+b}{2},\frac{d_{\rm (gauss)}\hat\omega_0^2}{\Lambda^2})}
         {\Gamma(\frac{b}{2})
     - \Gamma(\frac{b}{2},\frac{d_{\rm (gauss)}\hat\omega_0^2}{\Lambda^2})}
    \,, \nonumber\\
   \mu_\pi^2(\mu_f,\mu_i)
   &=& 3 \left[ \frac{\Lambda^2}{d_{\rm (gauss)}}\, 
    \frac{\Gamma(1+\frac{b}{2})
     - \Gamma(1+\frac{b}{2},\frac{d_{\rm (gauss)}\hat\omega_0^2}{\Lambda^2})}
    {\Gamma(\frac{b}{2})
     - \Gamma(\frac{b}{2},\frac{d_{\rm (gauss)}\hat\omega_0^2}{\Lambda^2})}
    - \bar\Lambda(\mu_f,\mu_i)^2 \right] .
\end{eqnarray}
The corresponding relations for $F^{\rm (hyp)}(\hat\omega;\Lambda,b)$ must 
be obtained numerically.

Ultimately the shape function should be fitted to the $\bar B\to X_s\gamma$
photon spectrum,
and the above equations then determine $\bar\Lambda$ and $\mu_\pi^2$. On
the other hand, these formulae can be inverted to determine
$\Lambda$ and $b$ from the current values of the HQET parameters. For
example, if we adopt the values $m_b(\mu_*,\mu_*)=4.61$\,GeV 
and $\mu_\pi^2(\mu_*,\mu_*)=0.20$\,GeV$^2$ for the parameters in 
(\ref{eq:atMuStar}) at $\mu_*=1.5$\,GeV, then we find the
parameter pair $\Lambda\approx 0.72$\,GeV, $b\approx 3.95$ for the
exponential model, $\Lambda\approx 0.71$\,GeV, $b\approx 2.36$ for
the gaussian model, and $\Lambda\approx 0.73$\,GeV, $b\approx 3.81$
for the hyperbolic model. On the right-hand side of
Figure~\ref{fig:SFmodels} we show these three different functions
plotted on the interval $[0,\hat\omega_0]$ over which the moment 
constraints are imposed. While the exponential
(solid) and hyperbolic (dash-dotted) curves are barely
distinguishable, the gaussian model has quite different
characteristics. It is broader, steeper at the onset, faster to
fall off, and the maximum is shifted toward larger $\hat\omega$. 

An important comment is that, 
once a two-parameter ansatz is employed, the shape-function parameters (i.e., 
$m_b$ and $\mu_\pi^2$) can either be determined from a fit to the entire 
photon spectrum, or to the first two moments of the spectrum. Both 
methods are equivalent and should yield consistent results. If they do not, 
it would be necessary to refine the ansatz for the functional form of the 
shape function. 

In most applications shape functions are needed for arguments
$\hat\omega$ of order $\Lambda_{\rm QCD}$. However, in some cases, like
the ideal cut on hadronic invariant mass, $\hat\omega$ is required to
be as large as $M_D$, which is much larger than $\Lambda_{\rm QCD}$. 
The large-$\hat\omega$ behavior of the shape functions can be computed
in a model-independent way using short-distance methods. 
For the leading shape function, one finds \cite{Bosch:2004th}
\begin{equation}\label{SFtail}
   \hat S(\hat\omega\gg\Lambda_{\rm QCD},\mu_i)
   = - \frac{C_F\alpha_s(\mu_i)}{\pi} \frac{1}{\hat\omega - \bar\Lambda} 
   \left( 2\ln\frac{\hat\omega-\bar\Lambda}{\mu_i} + 1 \right) 
   + \dots \,.
\end{equation}
Note that this radiative tail is negative, implying that 
the shape function must go through zero somewhere near $\hat\omega
\sim \mbox{few}\,\Lambda_{\rm QCD}$. For practical purposes, we ``glue'' the 
above expression onto models of the non-perturbative shape function starting
at $\hat\omega=\bar\Lambda+\mu_i/\sqrt{e}\approx 1.6$\,GeV, where the tail 
piece vanishes. In this way we obtain a continuous shape-function model with 
the correct asymptotic behavior. We stress that for applications with a 
maximal $P_+$ not larger than about 1.6\,GeV the radiative tail of the shape 
function is never required. This includes all methods for extracting 
$|V_{ub}|$ discussed later in this work, except for the case of a cut on 
hadronic invariant mass, $M_X\le M_0$, if $M_0$ is above 1.6\,GeV.

\subsection{Models for subleading shape functions}
\label{sec:modelSSF}

In the last section we have been guided by the fact that the 
$\bar B\to X_s\gamma$ photon spectrum is at leading power directly determined 
by the leading shape function. This helped in finding
models that have roughly the same shape as the photon spectrum. At the
subleading level considered here, however, no such guidance is
provided to us. The available information is limited to the tree-level
moment relations (\ref{SSF:moments}), stating that the norms of the subleading 
shape functions vanish while their first moments do not. 
In \cite{Bosch:2004cb}, two classes of models have been proposed, in which
the subleading shape functions are ``derived'' from the
leading shape function. A particularly simple choice is
\begin{equation}\label{SSF:model1}
   \hat t(\hat\omega) = - \lambda_2\,\hat S'(\hat\omega) \,, \qquad
   \hat u(\hat\omega) = \frac{2\lambda_1}{3}\,\hat S'(\hat\omega) \,, \qquad
   \hat v(\hat\omega) = \lambda_2\,\hat S'(\hat\omega) \,.
\end{equation}
Below, we will sometimes refer to this set of functions as the ``default
choice''. We choose the parameter $-\lambda_1$ in the expression for 
$\hat u(\hat\omega)$  (as well as in the expressions for the second-order 
hadronic power corrections) to coincide with the quantity 
$\mu_\pi^2(\mu_f,\mu_i)$ 
given in (\ref{parsExp}) and (\ref{parsGauss}). However, for consistency with 
the tree-level moment relations, we identity the parameter $\bar\Lambda$ in 
(\ref{BsgSSF2}) and (\ref{upsy}) with the quantity $\bar\Lambda(\mu_f,\mu_i)$ 
evaluated in the limit where $\omega_0\to\infty$. This implies 
$\bar\Lambda=\Lambda$ for all three types of functions and ensures that the 
subleading shape functions have zero norm when integrated over 
$0\le\hat\omega<\infty$. 

There are of course infinitely many possibilities to find models for
subleading shape functions that are in accordance with (\ref{SSF:moments}). 
Any function with vanishing norm and first moment can be arbitrarily added to 
any model for a subleading shape function without violating the moment 
relations. Several such functions have been proposed in recent work on 
subleading shape functions, see e.g.\
\cite{Bosch:2004cb,Neubert:2002yx,Beneke:2004in,Neubert:2004cu}.
Specifically, we define the functions
\begin{eqnarray} \label{hfuncs}
   h_1(\hat\omega)
   &=& \frac{M_2}{N\Omega_0^3}\,\frac{a^{a+1}}{2\Gamma(a)}\,z^{a-1}\,e^{-az}
    \left( \frac{a-1}{z} - a(2-z) \right) , \nonumber\\
   h_2(\hat\omega)
   &=& \frac{M_2}{N\Omega_0^3}\,\frac{a^3}{2}\,e^{-az}
    \left( 1 - 2az + \frac{a^2 z^2}{2} \right) , \nonumber \\ 
   h_3(\hat\omega)
   &=& \frac{M_2}{N\Omega_0^3} \left\{
    \left[ \frac{2\sqrt{\pi\,a}}{\pi-2}\,e^{-a z^2} 
    \left( 1 - 2z\sqrt{\frac a\pi} \right) \right]
    - 2e^{-z} + 2z\,e^{-2z}\,\mbox{Ei}\,(z) \right\} , \nonumber\\
   h_4(\hat\omega)
   &=& \frac{M_2}{N\Omega_0^3} \left\{
    \left[ \frac{\pi^2}{4}\,\frac{2\sqrt{\pi\,a}}{\pi-2}\,e^{-a z^2}
    \left( 1 - 2z\sqrt{\frac a\pi} \right) \right] \right. \nonumber\\
   &&\hspace{1.5cm} \left.
    + \frac 8 {\left(1+z^2\right)^4}\,
    \bigg[ z\ln z+ \frac z2\,(1+z^2) - \frac{\pi}{4}\,
    (1-z^2) \bigg] \right\} ,
\end{eqnarray}
where $z=\hat\omega/\Omega_0$, and the reference quantity 
$\Omega_0=O(\Lambda_{\rm QCD})$ depends on the type of function, namely 
$\Omega_0=\bar\Lambda$ for $h_1$ and $h_2$, $\Omega_0=\frac23\,\bar\Lambda$
for $h_3$, and $\Omega_0=\frac{8}{3\pi}\,\bar\Lambda$ for 
$h_4$. The quantity $a$ is a free parameter. The functions 
(\ref{hfuncs}) have by construction vanishing norm and first moment. 
Their second moments are given by the parameter $M_2$, provided the 
normalization constants are chosen as $N=1$ for $h_1$ and $h_2$, and 
\begin{equation}
   N = 1 - \frac{4-\pi}{2(\pi-2)}\,\frac{1}{a} \,, \qquad
   N = 1 - \frac{\pi^2(4-\pi)}{8(\pi-2)}\,\frac{1}{a}
\end{equation}
for  $h_3$ and  $h_4$, respectively. The values for the 
parameters $a$ and $M_2$ should be chosen such that the following 
characteristics of subleading shape functions are respected: First, they
are dimensionless functions, so that their values are naturally of
$O(1)$ for $\hat\omega\sim\Lambda_{\rm QCD}$. Secondly, when integrated 
over a sufficiently large domain, their contributions are
determined in terms of their first few moments. In particular, this 
implies that for values of $\hat\omega\gg\Lambda_{\rm QCD}$
the integrals over the subleading shape functions must approach zero.
Taking these considerations into account, we use $M_2=(0.3\,\mbox{GeV})^3$ 
in all cases and choose $a=3.5$ for $h_1$, $a=5$ for $h_2$, and $a=10$ for 
$h_3$ and $h_4$.

\begin{figure}[t]
\begin{center}
\epsfig{file=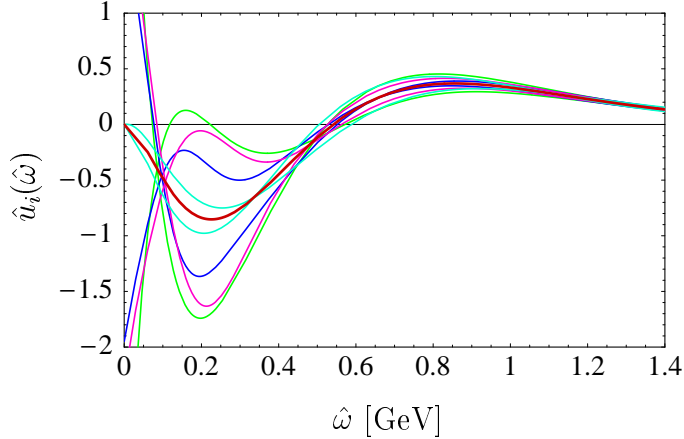, width=9cm}
\caption{\label{fig:SSFmodels}
Nine models for the subleading shape function $\hat u(\hat\omega)$ obtained 
by adding or subtracting one of the four functions $h_n(\hat\omega)$ to the 
default model in (\ref{SSF:model1}), shown as a thick line. See text for 
explanation.}
\end{center}
\end{figure}

Given the four functions (\ref{hfuncs}), we can construct several new
models for the subleading shape functions $\hat t(\hat\omega)$, 
$\hat u(\hat\omega)$, and $\hat v(\hat\omega)$. For each function, we 
construct a set of 9 models by adding or subtracting any of the functions 
$h_n(\hat\omega)$ to the default choice in (\ref{SSF:model1}). Together, this 
method yields $9^3=729$ different sets 
$\{\hat t_i(\hat\omega), \hat u_j(\hat\omega), \hat v_k(\hat\omega)\}$ with
$i,j,k=1,\dots,9$. 
This large collection of functions will be used to estimate the
hadronic uncertainties in our predictions for partial decay rates. Note 
that for most of these sets we no longer have 
$\hat t_i(\hat\omega)=-\hat v_k(\hat\omega)$, which was an ``accidental'' 
feature of the default model (\ref{SSF:model1}). 
The fact that the two functions have equal (but opposite in sign) first 
moments does not imply that their higher moments should also be related to 
each other.

For the case of $\hat u(\hat\omega)$ the resulting functions are shown in 
Figure~\ref{fig:SSFmodels}, where we have used the exponential model 
(\ref{SF:threeModels}) with parameters $\Lambda=0.72$\,GeV and $b=3.95$
for the leading shape function. In the region
$\hat\omega \sim \Lambda_{\rm QCD}$ they differ dramatically from each
other, while the large $\hat\omega$ dependence is dominated by the
moment relations (\ref{SSF:moments}). 

\subsection{Illustrative studies}

We stressed several times that the calculation of the hadronic tensor is 
``optimized'' for the shape-function region of large $P_-$ and small $P_+$, 
while it can smoothly be extended over the entire phase space. The notions
``large $P_-$'' and ``small $P_+$'' are to be understood as the sizes of 
integration domains for $P_-$ and $P_+$. Only when 
the differential distributions are integrated over a sufficiently 
large region in phase space, global quark-hadron duality ensures that the 
partonic description used in the present work matches the true, hadronic 
distributions with good accuracy. A more ambitious goal would be to 
calculate the differential decay rate point by point in the $(P_+,P_-)$ plane.
This can be done invoking local quark-hadron duality, as long as there is a 
sufficiently large number of hadronic final states contributing to 
the rate at any given point in phase space. 

\begin{figure}[t]
\begin{center}
\begin{minipage}{7.8cm}
\epsfig{file=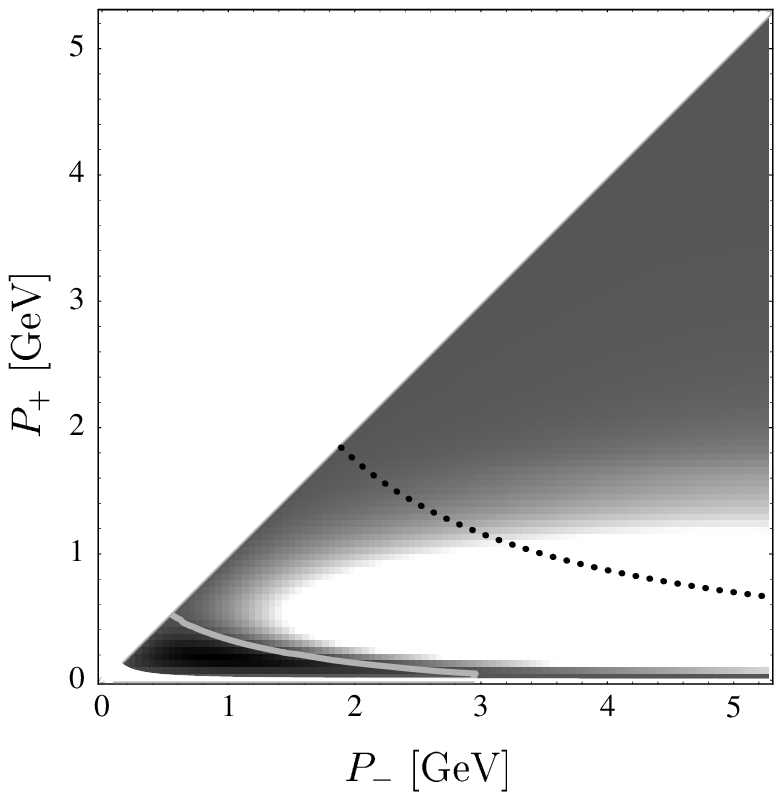, width=7.65cm}
\end{minipage}\hspace{5mm}
\begin{minipage}{8cm}
\epsfig{file=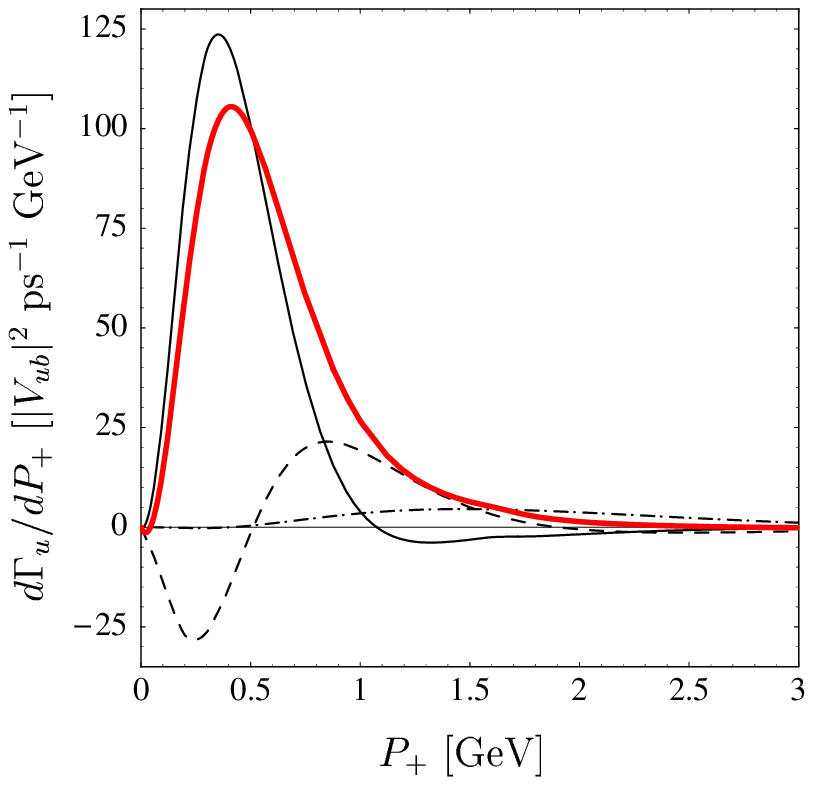, width=8.0cm}
\end{minipage}
\caption{\label{fig:d2Gamma}
{\sc Left:} Theoretical prediction for the double differential decay rate. 
The light area represents a large decay rate. Black regions
denote areas where the decay rate is close to zero. The dotted line is
given by $P_+ P_- = M_D^2$, which means that charm background is
located in the upper wedge. See text for further explanation.  
{\sc Right:} The $P_+$ spectrum extended to large values of $P_+$. The
thin solid line denotes the leading-power prediction, the dashed line
depicts first-order power corrections, the dash-dotted line shows 
second-order power corrections, and the thick solid line is their sum.}
\end{center}
\end{figure}

It is instructive to integrate the triple differential decay rate 
(\ref{eq:tripleRate}) over the leptonic variable $P_l$ in the range 
$P_+\le P_l\le P_-$, which yields the exact formula
\begin{eqnarray}\label{eq:doubleRate}
   \frac{d^2\Gamma_u}{dP_+\,dP_-}
   &=& \frac{G_F^2|V_{ub}|^2}{96\pi^3}\,U_y(\mu_h,\mu_i)\,
    (M_B-P_+)(P_- -P_+)^2 \nonumber\\
   &&\times \Big\{ (3M_B-2P_- -P_+)\,\F_1
    + 6 (M_B-P_-)\,\F_2 + (P_- -P_+)\,\F_3 \Big\} \,.
\end{eqnarray}
Our theoretical prediction for the double differential decay rate 
(\ref{eq:doubleRate}) is shown on the left-hand side of 
Figure~\ref{fig:d2Gamma}. We use the exponential model for the leading shape
function with parameters $m_b(\mu_*,\mu_*)=4.61$\,GeV and 
$\mu_\pi^2(\mu_*,\mu_*)=0.2$\,GeV$^2$, as well as the default choice 
(\ref{SSF:model1}) for the subleading shape functions. For very small 
$P_-$ values the rate turns negative (to the left of the gray 
line in the figure), signaling a breakdown of quark-hadron duality. 
It is reassuring that the only region where this happens is the 
``resonance region'', where the hadronic invariant mass is of order 
$\Lambda_{\rm QCD}$, and local duality breaks down.

Another useful quantity to consider is the differential $P_+$ rate,
which is obtained by integrating the double differential rate over $P_-$ in 
the range $P_+\le P_-\le M_B$. The resulting $P_+$ spectrum 
is shown on the right-hand side of Figure~\ref{fig:d2Gamma}. In the plot
we also disentangle the contributions from different orders in power
counting.

\section{Predictions and error estimates for partial rates}
\label{sec:analysio}

Before discussing predictions for partial $\bar B\to X_u\,l^-\bar\nu$ rates
for various kinematical cuts, let us recapitulate the ingredients of
the calculation and general procedure. We have presented expressions
for the triple differential decay rate, which can be organized in an expansion
in inverse powers of $(M_B-P_+)$. The leading-power contribution is given
at next-to-leading order in renormalization-group 
improved perturbation theory. At first
subleading power two contributions arise. The first type involves
subleading shape functions and is included at tree level, while the
second type contributes perturbative corrections of order $\alpha_s$ 
that come with the leading shape function. Further
contributions enter at second subleading power and are again of 
the two types: perturbative corrections of order $\alpha_s$ and
non-perturbative structures at tree level. In summary, then, partial rates 
can be computed term by term in an expansion of the form
\begin{equation}\label{partialRates}
   \Gamma_u = \Gamma_u^{\rm (0)}
   + (\Gamma_u^{\rm kin(1)} + \Gamma_u^{\rm hadr(1)})
   + (\Gamma_u^{\rm kin(2)} + \Gamma_u^{\rm hadr(2)}) + \dots \,.
\end{equation}
The goal of this section is to test the convergence of this series expansion 
and to 
perform a thorough analysis of uncertainties. For the kinematical corrections
$\Gamma_u^{{\rm kin}(n)}$ the sum of all terms is known and given by the 
expressions in (\ref{eq:fullDFN}), while the first two terms in the series 
correspond to the expanded results in (\ref{eq:kinNLO}) and 
(\ref{eq:kinNNLO}). We will find that in all cases of interest the first two 
terms give an excellent approximation to the exact result for 
$\Gamma_u^{\rm kin}$.

For the purpose of illustration, we adopt the exponential model for the shape 
function and present numerical results for two sets of input parameters, which
are biased by the results deduced from fits to $\bar B\to X_c\,l^-\bar\nu$ 
moments \cite{Neubert:2004sp}. Specifically, we use
$m_b(\mu_*,\mu_*)=4.61$\,GeV, $\mu_\pi^2(\mu_*,\mu_*)=0.2$\,GeV$^2$ (set~1) 
and $m_b(\mu_*,\mu_*)=4.55$\,GeV, $\mu_\pi^2(\mu_*,\mu_*)=0.3$\,GeV$^2$ 
(set~2). The values of the $b$-quark mass coincide with those obtained at 
two-loop and one-loop order in \cite{Neubert:2004sp} (see also the discussion 
below), while the values of $\mu_\pi^2$ are close to the corresponding values
in that reference. As was mentioned before, in the future 
the leading shape function $\hat\S(\hat\omega,\mu_i)$ 
should be extracted from a fit to the $\bar B\to X_s\gamma$ photon
spectrum, in which case the uncertainty in its shape becomes an experimental 
error, which can be systematically reduced with improved data. In the process, 
the ``theoretically preferred'' parameter values used in the present work 
will be replaced with the ``true'' values extracted directly from data.
While this will change the central values for the partial rates, our 
estimates of the theoretical errors will only be affected marginally. 

The different sources of theoretical uncertainties are as follows:
First, there are uncertainties associated with the functional forms
of the subleading shape functions. To estimate them, we take the spread of
results obtained when using the large set of different models described in 
Section~\ref{sec:modelSSF}, while the central value for a partial decay rate
corresponds to the default model (\ref{SSF:model1}). Secondly, there are 
perturbative uncertainties associated with the choice of the matching scales
$\mu_h$, $\mu_i$, and $\bar\mu$. Decay rates are formally independent
of these scales, but a residual dependence remains because of the truncation 
of the perturbative series. Our error analysis is as follows:
\begin{itemize}
\item 
The hard scale $\mu_h$ is of order $m_b$. In perturbative
logarithms the scale appears in the combination $(y m_b/\mu_h)$, see
e.g.\ (\ref{eq:LeadHardFunction}). To set a
central value for $\mu_h$ we are guided by the average $\langle
y\rangle m_b$. The leading term for the double differential decay rate
$d^2\Gamma_u /d P_+ dy$ is proportional to $2y^2(3-2y)$. It follows
that the average $y$ on the interval $[0,1]$ is 0.7. 
However, in some applications $y$ is not integrated over the full domain. 
Also, there are large negative constants in the matching correction $H_{u1}$ 
in (\ref{eq:LeadHardFunction}), whose effect can be ameliorated by lowering
the scale further. In the error analysis we use the central value 
of $\mu_h=m_b/2\approx 2.3$\,GeV and vary
the scale by a factor between $1/\sqrt2$ and $\sqrt2$. For the central value 
$\alpha_s(\mu_h)\approx 0.286$.
\item 
The intermediate scale $\mu_i\sim\sqrt{m_b \Lambda_{\rm QCD}}$ serves as the 
renormalization point for the jet and shape functions. 
We fix this scale to $\mu_i=1.5$\,GeV. Variations of $\mu_i$ would
affect both the normalization and the functional form of the shape function, 
as determined by the solution to the renormalization-group equation for the
shape function discussed in \cite{Bosch:2004th,Neubert:2004dd}. In practice, 
effects on the shape are irrelevant because the shape function is fitted to
data. The only place where the intermediate scale
has a direct impact on the extraction of $|V_{ub}|$ is through the
normalization of the shape function (\ref{SFnorm}). 
In the analysis we therefore estimate the uncertainty by assigning the
value $\pm (\frac{\alpha_s(\mu_i)}{\pi})^2$ as a relative error, where
$\alpha_s(\mu_i)\approx 0.354$. 
\item 
The scale $\bar\mu$ appears as the argument of $\alpha_s$ in the
perturbative contributions $\Gamma_u^{\rm kin}$. We vary $\bar\mu$ from 
$\mu_i/\sqrt{2}$ to $\sqrt{2}\mu_i$ with the central value 
$\bar\mu=\mu_i=1.5$\,GeV.
\end{itemize}
These three errors are added in quadrature and assigned as the total 
perturbative uncertainty. Finally, we need to estimate the effects from 
higher-dimensional operators at third and higher-order in power counting.  
If the considered cut includes the region of phase space near the origin 
($P_+\sim P_-\sim\Lambda_{\rm QCD}$), then the dominant such 
contributions are weak annihilation effects, which we have discussed in 
Section~\ref{Sec:WA}. From the analysis in \cite{TomsThesis} one can derive 
a bound on the weak annihilation contribution that is $\pm$1.8\% of the total 
decay rate, for which we take $\Gamma_u\approx 70\,|V_{ub}|^2\,{\rm ps}^{-1}$ 
(see below). The resulting uncertainty 
$\delta\Gamma_u^{\rm WA}=\pm 1.3\,|V_{ub}|^2\,{\rm ps}^{-1}$ affects all 
partial rates which include the region near the origin in the $(P_+,P_-)$ 
plane. The uncertainty from weak annihilation can be avoided by
imposing a cut $q^2\le q_{\rm max}^2$ (see Section~\ref{sec:eliminateWA}).
For all observables considered in the present work, other 
power corrections of order $1/m_b^3$ 
can be safely neglected. This can be seen by multiplying the contributions
from second-order hadronic power corrections to the various 
decay rates (called $\Gamma_u^{\rm hadr(2)}$) by an additional suppression 
factor $\Lambda_{\rm QCD}/m_b\sim 0.1$.

The following subsection contains a discussion of the total decay rate. In the
remainder of this section we then present predictions for a variety of 
kinematical cuts designed to eliminate (or reduce) the charm background. 
These partial rates can be computed either numerically or, in many cases, 
semi-analytically. In Appendix~\ref{appendix:B} we discuss how to perform the
integrations over the kinematical variables $P_l$ and $P_-$ analytically.

\subsection{Total decay rate}

Before presenting our predictions for the various partial decay rates, it
is useful to have an expression for the total $\bar B\to X_u\,l^-\bar\nu$ 
decay rate expressed in terms of the heavy-quark parameters defined in the 
shape-function scheme. We start from the exact two-loop expression for the
total rate derived in \cite{vanRitbergen:1999gs}, add the second-order 
hadronic power corrections, which are known at tree level 
\cite{Blok:1993va,Manohar:1993qn}, 
and finally convert the parameters $m_b$ and $\lambda_1$ from the pole scheme
to the shape-function scheme. The relevant replacements at two-loop order
can be taken from \cite{Neubert:2004sp} and read
\begin{eqnarray}
   m_b^{\rm pole} &=& m_b + 0.424\mu_*\alpha_s(\mu) \left[
    1 + \left( 1.357 + 1.326\ln\frac{\mu}{\mu_*}
    + 0.182\,\frac{\mu_\pi^2}{\mu_*^2} \right) \alpha_s(\mu) \right]
   \nonumber\\
   &&\mbox{}+ \frac{3\lambda_2-\mu_\pi^2-0.330\mu_*^2\alpha_s^2(\mu)}{2m_b}
    + \dots \,, \nonumber\\
   -\lambda_1 &=& \mu_\pi^2 + 0.330\mu_*^2\alpha_s^2(\mu) + \dots \,,
\end{eqnarray}
where here and from now on $m_b\equiv m_b(\mu_*,\mu_*)$ and
$\mu_\pi^2\equiv\mu_\pi^2(\mu_*,\mu_*)$ are defined in the shape-function 
scheme. At a reference scale $\mu_*=1.5$\,GeV the values of these 
parameters have been determined to be $m_b=(4.61\pm 0.08)$\,GeV and 
$\mu_\pi^2=(0.15\pm 0.07)$\,GeV$^2$ \cite{Neubert:2004sp},\footnote{The values 
obtained from a one-loop analysis are $m_b=(4.55\pm 0.08)$\,GeV and 
$\mu_\pi^2=(0.34\pm 0.07)$\,GeV$^2$.} 
where we account for the small $1/m_b$ correction to 
the relation for the pole mass in the above formula (corresponding to a 
shift of about $-0.02$\,GeV in $m_b$), which was not included in that paper.

The resulting expression for the total decay rate is
\begin{eqnarray}\label{sonice}
   \Gamma_u
   &=& \frac{G_F^2 |V_{ub}|^2 m_b^5}{192\pi^3}\,\Bigg\{
    1 + \alpha_s(\mu) \left( -0.768 + 2.122\,\frac{\mu_*}{m_b} \right)
    \nonumber\\
   &&\mbox{}+ \alpha_s^2(\mu) \left[ - 2.158 + 1.019\ln\frac{m_b}{\mu}
    + \left( 1.249 + 2.814\ln\frac{\mu}{\mu_*}
    + 0.386\,\frac{\mu_\pi^2}{\mu_*^2} \right) \frac{\mu_*}{m_b}
    + 0.811\,\frac{\mu_*^2}{m_b^2} \right] \nonumber\\
   &&\mbox{}- \frac{3(\mu_\pi^2-\lambda_2)}{m_b^2} + \dots \Bigg\} \,.
\end{eqnarray}
We observe that for $\mu_*\approx 1.5$\,GeV and $\mu=O(m_b)$, the perturbative 
expansion coefficients are strongly reduced compared to their values in the 
pole scheme ($-0.768$ and $-2.158$, respectively), indicating a vastly 
improved convergence of the perturbative expansion. For  
$m_b=\mu=4.61$\,GeV, and $\mu_\pi^2=0.15$\,GeV$^2$ we obtain for the 
one-loop, two-loop, and power corrections inside the brackets in 
(\ref{sonice}): $\{1 - 0.017 - 0.030 - 0.004\}$. All of these are very small
corrections to the leading term.

Including the uncertainties in the values of $m_b$ and $\mu_\pi^2$ quoted
above, and varying the renormalization scale $\mu$ between $m_b$ and $m_b/2$ 
(with a central value of $m_b/\sqrt2$), we get
\begin{equation}\label{totrate}
   \frac{\Gamma_u}{|V_{ub}|^2\,\mbox{ps}^{-1}}
   = 68.0_{\,-5.5}^{\,+5.9}\,[m_b]\,\mp 0.7\,[\mu_\pi^2]\,
   {}_{\,-0.9}^{\,+0.6}\,[\mu] 
   = \left( 68.0\mp 0.7\,[\mu_\pi^2]\,{}_{\,-0.9}^{\,+0.6}\,[\mu] \right)
   \left( \frac{m_b}{4.61\,\mbox{GeV}} \right)^{4.81} .
\end{equation}
Here and below, we quote values for decay rates in units of 
$|V_{ub}|^2\,{\rm ps}^{-1}$. To convert these results to partial branching 
fractions the numbers need to be multiplied by the average $B$-meson lifetime. 
Without including the two-loop corrections, the central value in the above 
estimate increases to 70.6. For comparison, with the same set of input 
parameters our new approach based on (\ref{eq:tripleRate}) predicts a total 
decay rate of 
$\Gamma_u=(71.4_{\,-5.0}^{\,+6.2}\pm 0.5)\,|V_{ub}|^2\,\mbox{ps}^{-1}$, where 
the first error accounts for perturbative uncertainties while the second one 
refers to the modeling of subleading shape functions (to which there is
essentially no sensitivity at all in the total rate). The fact that this 
is in excellent agreement with the direct calculation using (\ref{sonice}) 
supports the notion that the formalism developed in this work can be used to 
describe arbitrary $\bar B\to X_u\,l^-\bar\nu$ decay distributions, both in 
the shape-function region and in the OPE region of phase space.

\subsection{Cut on charged-lepton energy}

\begin{table}[t]
\begin{center}
\caption{\label{tab:LeptoInB}
Partial decay rate $\Gamma_u(E_0)$ for a cut on charged-lepton energy 
$E_l>E_0$ in the $B$-meson rest frame, given in units of 
$|V_{ub}|^2\,\mbox{ps}^{-1}$. Predictions are based on the shape-function 
parameter values $m_b=4.61$\,GeV, $\mu_\pi^2=0.2$\,GeV$^2$ (top) and 
$m_b=4.55$\,GeV, $\mu_\pi^2=0.3$\,GeV$^2$ (bottom).} 
\vspace{5mm}
\begin{tabular}{|c|c|c|c|c|} 
\hline
$E_0$ [GeV] & Mean & Subl.\ SF & Pert.\ & Total \\ 
\hline\hline
1.9 & 24.79 & $\pm$0.53 & $^{+1.90}_{-1.66}$ & $^{+2.34}_{-2.15}$ \\
2.0 & 18.92 & $\pm$0.60 & $^{+1.35}_{-1.20}$ & $^{+1.95}_{-1.84}$ \\
2.1 & 13.07 & $\pm$0.71 & $^{+0.82}_{-0.75}$ & $^{+1.66}_{-1.63}$ \\
2.2 &  7.59 & $\pm$0.81 & $^{+0.38}_{-0.34}$ & $^{+1.55}_{-1.54}$ \\
2.3 &  3.12 & $\pm$0.89 & $^{+0.15}_{-0.16}$ & $^{+1.55}_{-1.55}$ \\
2.4 &  0.42 & $\pm$1.05 & $^{+0.16}_{-0.22}$ & $^{+1.65}_{-1.65}$ \\
\hline
\hline
1.9 & 21.10 & $\pm$0.53 & $^{+1.57}_{-1.35}$ & $^{+2.08}_{-1.92}$ \\
2.0 & 15.83 & $\pm$0.60 & $^{+1.08}_{-0.94}$ & $^{+1.77}_{-1.68}$ \\
2.1 & 10.73 & $\pm$0.68 & $^{+0.64}_{-0.55}$ & $^{+1.57}_{-1.54}$ \\
2.2 &  6.12 & $\pm$0.74 & $^{+0.31}_{-0.23}$ & $^{+1.50}_{-1.48}$ \\
2.3 &  2.47 & $\pm$0.84 & $^{+0.17}_{-0.22}$ & $^{+1.53}_{-1.53}$ \\
2.4 &  0.29 & $\pm$0.99 & $^{+0.18}_{-0.24}$ & $^{+1.61}_{-1.62}$ \\
\hline
\end{tabular}
\end{center}
\end{table}

Traditionally, the most common variable to discriminate against the charm 
background is the charged-lepton energy $E_l$. As long as one requires that 
$E_l$ is bigger than $(M_B^2-M_D^2)/2M_B\approx 2.31$\,GeV, the final hadronic 
state cannot have an invariant mass larger than $M_D$. For this ideal cut, and 
using the default set of subleading shape functions, we find
\begin{equation}
\begin{array}{lllllll}
   & \Gamma_u^{\rm (0)}
   & +\,\, (\Gamma_u^{\rm kin(1)} & +\,\, \Gamma_u^{\rm hadr(1)}) 
   & +\,\, (\Gamma_u^{\rm kin(2)} & +\,\, \Gamma_u^{\rm hadr(2)}) & \\
   = \big[ & 6.810 & +\,\, (0.444 & -\,\, 3.967)
   & +\,\, (0.042 & -\,\, 0.555) \big] 
   & |V_{ub}|^2\,{\rm ps}^{-1} \,.
\end{array}
\end{equation}
The corrections from subleading shape functions are quite sizable, in 
accordance with the findings in
\cite{Bauer:2002yu,Leibovich:2002ys,Neubert:2002yx}. Note that 
the sum $\Gamma_u^{\rm kin(1)}+\Gamma_u^{\rm kin(2)}=0.486$ is an excellent
approximation to the exact result $\Gamma_u^{\rm kin}=0.482$ (all 
in units of $|V_{ub}|^2\,{\rm ps}^{-1}$) obtained
using (\ref{eq:fullDFN}), indicating that the expansion of the 
kinematical power corrections is converging rapidly. The same will be true
for all other observables considered below.

In practice, the cut on $E_l$ can be relaxed to some extent because the
background is well understood, thereby increasing the efficiency and reducing
the impact of theoretical uncertainties.
Our findings for different values of the cut $E_0$ are summarized in 
Table~\ref{tab:LeptoInB}. Here and below, the columns have the following 
meaning: ``Mean'' denotes the
prediction for the partial decay rate, ``Subl.\ SF'' the
uncertainty from subleading shape functions, and ``Pert.'' the total
perturbative uncertainty. In the column ``Total'' we add the
stated errors plus the uncertainty from weak annihilation in quadrature. 

\begin{table}[t]
\begin{center}
\caption{\label{tab:LeptoInUps}
Same as Table~\ref{tab:LeptoInB}, but for the partial decay rate 
$\gamma\,\Gamma_u^{(\Upsilon)}(E_0)$ for a cut on lepton energy $E_l>E_0$ in 
the $\Upsilon(4S)$ rest frame.} 
\vspace{5mm}
\begin{tabular}{|c|c|c|c|c|} 
\hline
$E_0$ [GeV] & Mean & Subl.\ SF & Pert.\ & Total \\ 
\hline\hline
1.9 & 24.82 & $\pm$0.54 & $^{+1.91}_{-1.66}$ & $^{+2.35}_{-2.15}$ \\
2.0 & 19.00 & $\pm$0.61 & $^{+1.37}_{-1.21}$ & $^{+1.96}_{-1.85}$ \\
2.1 & 13.25 & $\pm$0.71 & $^{+0.85}_{-0.76}$ & $^{+1.68}_{-1.63}$ \\
2.2 &  7.99 & $\pm$0.78 & $^{+0.42}_{-0.37}$ & $^{+1.54}_{-1.53}$ \\
2.3 &  3.83 & $\pm$0.86 & $^{+0.18}_{-0.13}$ & $^{+1.54}_{-1.53}$ \\
2.4 &  1.31 & $\pm$0.99 & $^{+0.10}_{-0.14}$ & $^{+1.61}_{-1.61}$ \\
\hline
\hline
1.9 & 21.16 & $\pm$0.54 & $^{+1.58}_{-1.35}$ & $^{+2.09}_{-1.93}$ \\
2.0 & 15.94 & $\pm$0.60 & $^{+1.10}_{-0.95}$ & $^{+1.78}_{-1.69}$ \\
2.1 & 10.94 & $\pm$0.68 & $^{+0.66}_{-0.57}$ & $^{+1.58}_{-1.54}$ \\
2.2 &  6.49 & $\pm$0.74 & $^{+0.34}_{-0.26}$ & $^{+1.50}_{-1.48}$ \\
2.3 &  3.05 & $\pm$0.84 & $^{+0.17}_{-0.18}$ & $^{+1.53}_{-1.53}$ \\
2.4 &  0.98 & $\pm$0.92 & $^{+0.13}_{-0.18}$ & $^{+1.56}_{-1.57}$ \\
\hline
\end{tabular}
\end{center}
\end{table}

Experiments often do not measure the partial rates in the $B$-meson
rest frame, but in the rest frame of the $\Upsilon(4S)$ resonance produced 
in $e^+ e^-$ collisions. Boosting to the
$\Upsilon(4S)$ frame with $\beta=v/c\approx 0.064$ has a small
effect on the spectrum and rates. The exact formula for this boost is 
\cite{Kagan:1998ym}
\begin{equation}
   \gamma\,\Gamma_u^{(\Upsilon)}(E_0) = \frac{1}{\beta_+ -\beta_-} 
   \int_{\beta_- E_0}^{M_B/2}\!dE\,\frac{d\Gamma_u^{(B)}}{dE} 
   \left[ \beta_+ - \mbox{max}\left( \beta_-,\frac{E_0}{E} \right) 
   \right] ,
\end{equation}
where $\beta_\pm=\sqrt{1\pm\beta}/\sqrt{1\mp\beta}$, and the factor
$\gamma=1/\sqrt{1-\beta^2}\approx 1.002$ on the left-hand side takes the 
time dilation of 
the $B$-meson lifetime $\tau_B'=\gamma\,\tau_B$ into account. (In other
words, branching fractions are Lorentz invariant.) 
The above formula can be accurately
approximated by the first term in an expansion in $\beta^2$, which yields
\cite{Kagan:1998ym}
\begin{equation}
   \gamma\,\Gamma_u^{(\Upsilon)}(E_0) = \Gamma_u^{(B)}(E_0) 
   - \frac{\beta^2}{6} E_0^3 
   \left[ \frac{d}{dE}\,\frac{1}{E}\,\frac{d\Gamma_u^{(B)}}{dE}
   \right]_{E=E_0} + O(\beta^4) \,,
\end{equation}
as long as $E_0$ is not too close to the kinematical endpoint (i.e., 
$E_0\le\beta_- M_B/2\approx 2.47$\,GeV).  The
numerical results for the partial decay rate
$\gamma\,\Gamma_u^{(\Upsilon)}(E_0)$ in the rest frame of the $\Upsilon(4S)$
resonance are given in Table~\ref{tab:LeptoInUps}. 

\subsection{Cut on hadronic {\boldmath $P_+$}}

Cutting on $P_+$ samples the same hadronic phase space as a cut on the
charged-lepton energy, but with much better efficiency 
\cite{Bosch:2004th,Bosch:2004bt}. The phase space $P_+\le\Delta_P$ with the 
ideal separator $\Delta_P=M_D^2/M_B\approx 0.66$\,GeV contains well over half 
of all $\bar B\to X_u\,l^-\bar\nu$ events. Here we find with the default 
settings
\begin{equation}
\begin{array}{lllllll}
   & \Gamma_u^{\rm (0)}
   & +\,\, (\Gamma_u^{\rm kin(1)} & +\,\, \Gamma_u^{\rm hadr(1)}) 
   & +\,\, (\Gamma_u^{\rm kin(2)} & +\,\, \Gamma_u^{\rm hadr(2)}) & \\
   = \big[ & 53.225 & +\,\, (4.646 & -\,\, 11.862)
   & +\,\, (0.328 & -\,\, 0.227) \big] 
   & |V_{ub}|^2\,{\rm ps}^{-1} \,.
\end{array}
\end{equation}
We see a much better convergence of the power series than in the case of a 
cut on the charged-lepton energy, namely $53.225-7.216-0.100$ when
grouping the above numbers according to their power counting. Once again, 
the sum $\Gamma_u^{\rm kin(1)}+\Gamma_u^{\rm kin(2)}=4.973$ 
is very close to the full kinematical correction 
$\Gamma_u^{\rm kin}=4.959$ (in units of $|V_{ub}|^2\,{\rm ps}^{-1}$).

\begin{table}[p]
\begin{center}
\caption{\label{tab:Pp}
Partial decay rate $\Gamma_u(\Delta_P, E_0)$ for a cut on the hadronic 
variable $P_+\le\Delta_P$ and lepton energy $E_l \ge E_0$, given in units of 
$|V_{ub}|^2\,\mbox{ps}^{-1}$. Predictions are based on the shape-function 
parameter values $m_b=4.61$\,GeV, $\mu_\pi^2=0.2$\,GeV$^2$ (top) and 
$m_b=4.55$\,GeV, $\mu_\pi^2=0.3$\,GeV$^2$ (bottom).} 
\vspace{5mm}
\begin{tabular}{|c|c|c|c|c|c|} 
\hline
$\Delta_P$ [GeV] & $E_0$ [GeV] & Mean & Subl.\ SF & Pert.\ & Total \\ 
\hline\hline
0.70 & 0.0 & 48.90 & $\pm$1.15 & $^{+2.83}_{-2.65}$ & $^{+3.30}_{-3.15}$ \\
0.65 & 0.0 & 45.34 & $\pm$1.46 & $^{+2.55}_{-2.41}$ & $^{+3.20}_{-3.09}$ \\
0.60 & 0.0 & 41.34 & $\pm$1.76 & $^{+2.26}_{-2.15}$ & $^{+3.13}_{-3.05}$ \\
0.55 & 0.0 & 36.91 & $\pm$2.01 & $^{+1.95}_{-1.87}$ & $^{+3.08}_{-3.02}$ \\
0.50 & 0.0 & 32.09 & $\pm$2.34 & $^{+1.64}_{-1.58}$ & $^{+3.12}_{-3.09}$ \\
\hline
0.70 & 1.0 & 43.36 & $\pm$1.02 & $^{+2.54}_{-2.39}$ & $^{+3.01}_{-2.88}$ \\
0.65 & 1.0 & 40.18 & $\pm$1.30 & $^{+2.28}_{-2.16}$ & $^{+2.92}_{-2.82}$ \\
0.60 & 1.0 & 36.59 & $\pm$1.59 & $^{+2.01}_{-1.92}$ & $^{+2.86}_{-2.80}$ \\
0.55 & 1.0 & 32.61 & $\pm$1.86 & $^{+1.73}_{-1.67}$ & $^{+2.84}_{-2.80}$ \\
0.50 & 1.0 & 28.29 & $\pm$2.19 & $^{+1.44}_{-1.40}$ & $^{+2.91}_{-2.89}$ \\
\hline
\hline
0.70 & 0.0 & 39.95 & $\pm$1.19 & $^{+2.18}_{-2.06}$ & $^{+2.79}_{-2.70}$ \\
0.65 & 0.0 & 36.94 & $\pm$1.42 & $^{+1.95}_{-1.86}$ & $^{+2.72}_{-2.66}$ \\
0.60 & 0.0 & 33.67 & $\pm$1.65 & $^{+1.71}_{-1.65}$ & $^{+2.69}_{-2.65}$ \\
0.55 & 0.0 & 30.15 & $\pm$1.88 & $^{+1.47}_{-1.43}$ & $^{+2.70}_{-2.68}$ \\
0.50 & 0.0 & 26.40 & $\pm$2.09 & $^{+1.22}_{-1.21}$ & $^{+2.73}_{-2.72}$ \\
\hline
0.70 & 1.0 & 35.42 & $\pm$1.13 & $^{+1.95}_{-1.85}$ & $^{+2.59}_{-2.51}$ \\
0.65 & 1.0 & 32.73 & $\pm$1.34 & $^{+1.74}_{-1.66}$ & $^{+2.53}_{-2.48}$ \\
0.60 & 1.0 & 29.81 & $\pm$1.55 & $^{+1.52}_{-1.47}$ & $^{+2.51}_{-2.48}$ \\
0.55 & 1.0 & 26.65 & $\pm$1.76 & $^{+1.29}_{-1.27}$ & $^{+2.52}_{-2.51}$ \\
0.50 & 1.0 & 23.29 & $\pm$1.95 & $^{+1.07}_{-1.06}$ & $^{+2.56}_{-2.55}$ \\
\hline
\end{tabular}
\end{center}
\end{table}

Often times it is required to impose an additional cut on the
charged-lepton energy, as leptons that are too soft are difficult to
detect. In Table~\ref{tab:Pp} we list results for both $E_l\ge 0$ and
$E_l\ge 1.0$ GeV. For the ideal cut we find that the prediction is
quite precise, as the total theoretical uncertainty is only about 6.8\%. For
comparison, the ideal cut for the lepton energy is uncertain by about
50\%, but rapidly improving as the energy cut is relaxed.

\subsection{Cut on hadronic invariant mass and {\boldmath $q^2$}}

\begin{table}[p]
\begin{center}
\caption{\label{tab:MxQ2}
Partial decay rate $\Gamma_u(M_0,q_0^2,E_0)$ for combined cuts 
$M_X\le M_0$ on hadronic invariant mass, $q^2>q_0^2$ on leptonic invariant 
mass, and $E_l\ge E_0$ on charged-lepton energy, given in units of 
$|V_{ub}|^2\,\mbox{ps}^{-1}$. Predictions are based on the shape-function 
parameter values $m_b=4.61$\,GeV, $\mu_\pi^2=0.2$\,GeV$^2$ (top) and 
$m_b=4.55$\,GeV, $\mu_\pi^2=0.3$\,GeV$^2$ (bottom).} 
\vspace{5mm}
\begin{tabular}{|c|c|c|c|c|c|c|} 
\hline
$M_0$ [GeV] & $q_0^2$ [GeV$^2$] & $E_0$ [GeV] & Mean & Subl.\ SF & Pert.\
 & Total \\
\hline\hline
$M_D$ & 0.0 & 0.0 & 59.30 & $\pm$0.36 & $^{+4.22}_{-3.73}$
 & $^{+4.42}_{-3.96}$ \\
1.70  & 0.0 & 0.0 & 53.13 & $\pm$0.73 & $^{+3.67}_{-3.31}$
 & $^{+3.95}_{-3.61}$ \\
1.55  & 0.0 & 0.0 & 45.72 & $\pm$1.16 & $^{+3.11}_{-2.84}$
 & $^{+3.55}_{-3.32}$ \\
$M_D$ & 6.0 & 0.0 & 34.37 & $\pm$0.37 & $^{+2.97}_{-2.58}$
 & $^{+3.25}_{-2.89}$ \\
1.70  & 8.0 & 0.0 & 24.80 & $\pm$0.36 & $^{+2.24}_{-1.98}$
 & $^{+2.59}_{-2.37}$ \\
$M_D$ & $(M_B-M_D)^2$ & 0.0 & 12.55 & $\pm$0.49 & $^{+1.41}_{-1.24}$
 & $^{+1.95}_{-1.83}$ \\
\hline
$M_D$ & 0.0 & 1.0 & 53.49 & $\pm$0.36 & $^{+3.91}_{-3.45}$
 & $^{+4.13}_{-3.69}$ \\
1.70  & 0.0 & 1.0 & 48.25 & $\pm$0.63 & $^{+3.42}_{-3.08}$
 & $^{+3.70}_{-3.38}$ \\
1.55  & 0.0 & 1.0 & 41.81 & $\pm$1.03 & $^{+2.91}_{-2.66}$
 & $^{+3.34}_{-3.12}$ \\
$M_D$ & 6.0 & 1.0 & 33.88 & $\pm$0.37 & $^{+2.94}_{-2.55}$
 & $^{+3.22}_{-2.87}$ \\
1.70  & 8.0 & 1.0 & 24.74 & $\pm$0.36 & $^{+2.23}_{-1.97}$
 & $^{+2.59}_{-2.37}$ \\
$M_D$ & $(M_B-M_D)^2$ & 1.0 & 12.55 & $\pm$0.49 & $^{+1.41}_{-1.24}$
 & $^{+1.95}_{-1.83}$ \\
\hline
\hline
$M_D$ & 0.0 & 0.0 & 50.08 & $\pm$0.54 & $^{+3.52}_{-3.11}$
 & $^{+3.78}_{-3.40}$ \\
1.70  & 0.0 & 0.0 & 44.20 & $\pm$0.86 & $^{+2.98}_{-2.69}$
 & $^{+3.35}_{-3.09}$ \\
1.55  & 0.0 & 0.0 & 37.76 & $\pm$1.22 & $^{+2.46}_{-2.26}$
 & $^{+3.03}_{-2.86}$ \\
$M_D$ & 6.0 & 0.0 & 29.42 & $\pm$0.35 & $^{+2.50}_{-2.16}$
 & $^{+2.82}_{-2.52}$ \\
1.70  & 8.0 & 0.0 & 20.87 & $\pm$0.39 & $^{+1.84}_{-1.61}$
 & $^{+2.26}_{-2.08}$ \\
$M_D$ & $(M_B-M_D)^2$ & 0.0 &10.49 & $\pm$0.48 & $^{+1.16}_{-1.00}$
 & $^{+1.76}_{-1.68}$ \\
\hline
$M_D$ & 0.0 & 1.0 & 45.29 & $\pm$0.50 & $^{+3.27}_{-2.88}$
 & $^{+3.54}_{-3.18}$ \\
1.70  & 0.0 & 1.0 & 40.22 & $\pm$0.77 & $^{+2.78}_{-2.50}$
 & $^{+3.15}_{-2.90}$ \\
1.55  & 0.0 & 1.0 & 34.55 & $\pm$1.09 & $^{+2.31}_{-2.11}$
 & $^{+2.85}_{-2.69}$ \\
$M_D$ & 6.0 & 1.0 & 28.99 & $\pm$0.34 & $^{+2.48}_{-2.13}$
 & $^{+2.80}_{-2.50}$ \\
1.70  & 8.0 & 1.0 & 20.82 & $\pm$0.39 & $^{+1.83}_{-1.60}$
 & $^{+2.26}_{-2.08}$ \\
$M_D$ & $(M_B-M_D)^2$ & 1.0 &10.49 & $\pm$0.48 & $^{+1.16}_{-1.00}$ &
 $^{+1.78}_{-1.68}$ \\
\hline
\end{tabular}
\end{center}
\end{table}

The most efficient separator for the discrimination of 
$\bar B\to X_c l^-\bar\nu$ events is a cut on the invariant mass $M_X$ of the 
hadronic final state, $M_X\le M_D$ \cite{Falk:1997gj,Bigi:1997dn}.
It has also been argued \cite{Bauer:2000xf} that a cut on 
$q^2$ can reduce the shape-function sensitivity, since it avoids the 
collinear region in phase space where $P_-\gg P_+$.
In order to optimize signal efficiency and theoretical uncertainties, it was
suggested in \cite{Bauer:2001rc} to combine a $q^2$ cut with a cut on hadronic 
invariant mass. 

The theoretical predictions obtained in \cite{Bauer:2000xf,Bauer:2001rc} 
were based on a conventional OPE calculation, which was assumed to be valid 
for these cuts. The
assessment of the shape-function sensitivity was based on convolving
the tree-level decay rate with a ``tree-level shape function'', for which 
two models (a realistic model similar to the ones considered here, and an 
unrealistic $\delta$-function model) were employed. The 
shape-function sensitivity was then inferred from the comparison of the 
results obtained with the two models. The sensitivity to subleading 
shape functions was not considered, since it was assumed to be very small.
Since our formalism smoothly interpolates between the
``shape-function'' and ``OPE'' regions, and since we include radiative 
corrections as well as power corrections as far as they are known, 
we can estimate 
the sensitivity of a combined $M_X$--$q^2$ cut to the leading and subleading 
shape functions much more accurately. Contrary to \cite{Bauer:2001rc}, 
we do not find a significant reduction of the shape-function sensitivity when 
adding the $q^2$ cut to a cut on hadronic invariant mass. 

In Table~\ref{tab:MxQ2} we give results for typical 
cuts on $M_X$ and $q^2$, with and without including an additional cut on 
charged-lepton energy.
Let us study the contributions for the optimal cut $M_X\le M_D$ in
detail. We find with the default settings
\begin{equation}
\begin{array}{lllllll}
   & \Gamma_u^{\rm (0)} 
   & +\,\, (\Gamma_u^{\rm kin(1)} & +\,\, \Gamma_u^{\rm hadr(1)}) 
   & +\,\, (\Gamma_u^{\rm kin(2)} & +\,\, \Gamma_u^{\rm hadr(2)}) & \\
   = \big[ & 58.541 & +\,\, (8.027 & -\,\, 9.048)
   & +\,\, (2.100 & -\,\, 0.318) \big] 
   & |V_{ub}|^2\,{\rm ps}^{-1} \,.
\end{array}
\end{equation}
Note the almost perfect (accidental) cancellation of the two terms at order 
$1/m_b$. The resulting power series,
$58.541-1.022+1.782$, again exhibits good convergence. As previously, the sum
$\Gamma_u^{\rm kin(1)}+\Gamma_u^{\rm kin(2)}=10.127$ is a good
approximation to the exact value 
$\Gamma_u^{\rm kin}=9.753$ (in units of $|V_{ub}|^2\,{\rm ps}^{-1}$). 
The analogous analysis for a combined cut $M_X\le 1.7$\,GeV and 
$q^2\ge 8.0$\,GeV$^2$ reads
\begin{equation}
\begin{array}{lllllll}
   & \Gamma_u^{\rm (0)} 
   & +\,\, (\Gamma_u^{\rm kin(1)} & +\,\, \Gamma_u^{\rm hadr(1)}) 
   & +\,\, (\Gamma_u^{\rm kin(2)} & +\,\, \Gamma_u^{\rm hadr(2)}) & \\
   = \big[ & 25.880 & +\,\, (4.049 & -\,\, 6.358)
   & +\,\, (1.399 & -\,\, 0.171) \big] 
   & |V_{ub}|^2\,{\rm ps}^{-1} \,,
\end{array}
\end{equation}
which means that the power series is $25.880-2.309+1.228$. Here we have 
$\Gamma_u^{\rm kin(1)}+\Gamma_u^{\rm kin(2)}=5.449$, which is close to
$\Gamma_u^{\rm kin}=5.160$ (in units of $|V_{ub}|^2\,{\rm ps}^{-1}$).

\subsection{Cut on {\boldmath $s_H^{\rm max}$} and {\boldmath $E_l$}}
\label{sec:babarcut}

In \cite{Aubert:2004tw}, the BaBar collaboration employed a
cut on both $E_l\ge E_0$ and a new kinematical variable 
$s_H^{\rm max}\le s_0$, where the
definition for $s_H^{\rm max}$ involves both hadronic and leptonic
variables. In the $B$-meson rest frame, it is
\begin{equation}\label{shmax}
   s_H^{\rm max} = M_B^2 + q^2 -2 M_B \left( E_l + \frac{q^2}{4E_l} \right) .
\end{equation}
Rewriting the phase space of this cut in the variables $P_+$, $P_-$, $P_l$, we 
find
\begin{eqnarray}
   0\le P_+ &\le& \mbox{min}\left( M_B-2E_0, \sqrt{s_0} \right) ,
    \nonumber \\
   P_+\le P_- &\le& \mbox{min}\left( \frac{s_0}{P_+}, M_B \right) ,
    \nonumber \\
   P_+\le P_l &\le& \mbox{min}\left( M_B-2E_0, P_- \right) ,
\end{eqnarray}
where it is understood that if $q^2=(M_B-P_+)(M_B-P_-)\le (M_B-\sqrt{s_0})^2$, 
then the interval $P_l^{\rm min}<P_l<P_l^{\rm max}$ must be {\em excluded\/} 
from the $P_l$ integration. Here
\begin{equation}
   P_l^{\rm max/min}(P_+,P_-) = \left(\frac{P_+ + P_-}{2}
   + \frac{s_0 - P_+ P_-}{2M_B} \right)
   \pm \sqrt{\left(\frac{P_+ + P_-}{2} + \frac{s_0 - P_+ P_-}{2M_B} \right)^2
   - s_0} \,.
\end{equation}

\begin{table}[t]
\begin{center}
\caption{\label{tab:sHmax}
Partial decay rate $\Gamma_u(s_0,E_0)$ for combined cuts 
$s_H^{\rm max}\le s_0$ and $E_l\ge E_0$, given in 
units of $|V_{ub}|^2\,\mbox{ps}^{-1}$. Predictions are based on the 
shape-function parameter values $m_b=4.61$\,GeV, $\mu_\pi^2=0.2$\,GeV$^2$ 
(top) and $m_b=4.55$\,GeV, $\mu_\pi^2=0.3$\,GeV$^2$ (bottom).} 
\vspace{5mm}
\begin{tabular}{|c|c|c|c|c|c|} 
\hline
$s_0$ [GeV$^2$] & $E_0$ [GeV] & Mean & Subl.\ SF & Pert.\ & Total \\ 
\hline\hline
3.5 & 1.8 & 17.39 & $\pm$0.62 & $^{+1.54}_{-1.36}$ & $^{+2.08}_{-1.96}$ \\ 
3.5 & 1.9 & 15.86 & $\pm$0.63 & $^{+1.33}_{-1.18}$ & $^{+1.94}_{-1.84}$ \\
3.5 & 2.0 & 13.70 & $\pm$0.66 & $^{+1.05}_{-0.94}$ & $^{+1.77}_{-1.71}$ \\
3.5 & 2.1 & 10.78 & $\pm$0.73 & $^{+0.71}_{-0.64}$ & $^{+1.62}_{-1.59}$ \\
\hline
\hline
3.5 & 1.8 & 14.57 & $\pm$0.60 & $^{+1.25}_{-1.09}$ & $^{+1.87}_{-1.77}$ \\
3.5 & 1.9 & 13.18 & $\pm$0.61 & $^{+1.06}_{-0.92}$ & $^{+1.76}_{-1.68}$ \\
3.5 & 2.0 & 11.28 & $\pm$0.64 & $^{+0.82}_{-0.71}$ & $^{+1.63}_{-1.58}$ \\
3.5 & 2.1 &  8.77 & $\pm$0.69 & $^{+0.54}_{-0.46}$ & $^{+1.54}_{-1.51}$ \\
\hline
\end{tabular}
\end{center}
\end{table}

A summary of our findings is given in Table~\ref{tab:sHmax}. When compared to 
the pure charged-lepton energy cut in Table~\ref{tab:LeptoInB}, 
the additional cut on $s_H^{\rm max}$
eliminates roughly another 20--30\% of events. However, the hope is
that this cut also reduces the sensitivity to the leading shape
function, which we expect to be sizable for the pure $E_l$ cut. The
uncertainty from subleading shape functions, however, is almost
unaffected by the $s_H^{\rm max}$ cut. 

\subsection{Eliminating weak annihilation contributions}
\label{sec:eliminateWA}

\begin{table}[t]
\begin{center}
\caption{\label{tab:withQ2}
Examples of partial decay rates with a cut on $q^2\le(M_B-M_D)^2$ imposed to 
eliminate the weak annihilation contribution. We consider an additional cut on 
the hadronic variable $P_+\le\Delta_P$ (top), or on the hadronic invariant 
mass $M_X\le M_0$ (bottom). As before, decay rates are given in units of 
$|V_{ub}|^2\,\mbox{ps}^{-1}$. Predictions are based on the shape-function 
parameters $m_b=4.61$\,GeV and $\mu_\pi^2=0.2$\,GeV$^2$.}
\vspace{5mm}
\begin{tabular}{|c|c|c|c|c|}
\hline
$\Delta_P$ [GeV] & Mean & Subl.\ SF & Pert.\ & Total \\ 
\hline\hline
0.70 & 39.96 & $\pm$1.27 & $^{+2.16}_{-2.01}$ & $^{+2.51}_{-2.38}$ \\ 
0.65 & 37.18 & $\pm$1.50 & $^{+1.99}_{-1.85}$ & $^{+2.49}_{-2.38}$ \\
0.60 & 34.05 & $\pm$1.71 & $^{+1.82}_{-1.69}$ & $^{+2.50}_{-2.41}$ \\
0.55 & 30.61 & $\pm$1.89 & $^{+1.63}_{-1.52}$ & $^{+2.49}_{-2.42}$ \\
0.50 & 26.86 & $\pm$1.97 & $^{+1.44}_{-1.33}$ & $^{+2.44}_{-2.38}$ \\
\hline\hline
$M_0$ [GeV] & Mean & Subl.\ SF & Pert.\ & Total \\ 
\hline\hline
$M_D$ & 46.75 & $\pm$0.65 & $^{+2.82}_{-2.50}$ & $^{+2.89}_{-2.58}$ \\ 
 1.70 & 40.70 & $\pm$1.12 & $^{+2.32}_{-2.11}$ & $^{+2.58}_{-2.39}$ \\ 
 1.55 & 33.69 & $\pm$1.56 & $^{+1.88}_{-1.73}$ & $^{+2.44}_{-2.32}$ \\
\hline
\end{tabular}
\end{center}
\end{table}

In Section~\ref{Sec:WA} we have argued that a cut on {\em high\/}
$q^2$, i.e., $q^2<q_0^2$, will eliminate the effect of weak
annihilation and remove the uncertainty associated with this
contribution. The cutoff $q_0^2$ should be small enough to exclude the
region around $q^2=m_b^2$, where this contribution is concentrated. It
is instructive to assess the ``cost'' of such an additional cut in
terms of the loss of efficiency and, more importantly, the behavior of
the remaining uncertainties. In order to do this, we combine the cut
$q^2\le(M_B-M_D)^2$ with either a cut on $P_+$ or on $M_X$. 
While this particular choice for $q_0^2$ still leaves some room
to improve the efficiency by increasing $q_0^2$, it is not desirable
to raise the cut much further, since this would threaten the validity
of quark-hadron duality.

The results are summarized in Table~\ref{tab:withQ2} and can be
compared to the previous ``pure'' $P_+$ and $M_X$ cuts in
Tables~\ref{tab:Pp} and \ref{tab:MxQ2}. As an example, let us consider
the case $P_+\le 0.65$ GeV, which is close to the charm
threshold. Without the additional $q^2$ cut we found that the total
theoretical uncertainty (including the weak annihilation error) is
${}_{-6.8}^{+7.0}$\%. When cutting in addition on $q^2\le(M_B-M_D)^2$,
the efficiency decreases by about 20\% as expected. However, due to the
absence of the weak annihilation uncertainty, the overall uncertainty
decreases to ${}_{-6.4}^{+6.7}$\%. Therefore both strategies result in
comparable relative uncertainties, with a slight favor for imposing
the additional cut from the theoretical point of view.

While the small reduction of theoretical errors hardly seems worth the effort 
of imposing the $q^2$ cut, performing an analysis of the type outlined 
here and comparing its results with those obtained without the additional cut 
may help to corroborate the expectation 
that the weak annihilation contribution is 
indeed not much larger than what has been found in \cite{TomsThesis}.

\subsection{Dependence on {\boldmath $m_b$} and shape-function sensitivity}

Non-perturbative hadronic physics enters in our approach via the form of the
leading and subleading shape functions. The strongest sensitivity by far is 
to the first moment of the leading shape function, which determines the HQET
parameter $\bar\Lambda$ and with it the $b$-quark mass. Given that the 
value of $m_b\equiv m_b(\mu_*,\mu_*)$ can be determined with good precision 
from other sources (such as moments of the leptonic or hadronic invariant 
mass spectra in $\bar B\to X_c\,l^-\bar\nu$ decays), it is instructive to 
disentangle this dependence from the sensitivity to 
higher moments or, more generally, to the functional form of the shape 
functions for fixed $m_b$. 

\begin{table}[t]
\begin{center}
\caption{\label{tab:mbdep} 
Values of the exponent $a(m_b)$ for different kinematical cuts. The parameter
$\mu_\pi^2=0.2$\,GeV$^2$ is kept fixed. Also quoted is the sensitivity of the 
partial decay rates to the functional form of the shape functions. See text 
for explanation.}
\vspace{5mm}
\begin{tabular}{|l|c|c|c|c|c|c|}
\hline
& $m_b$ [GeV] & 4.50 & 4.55 & 4.60 & 4.65 & 4.70 \\
\hline\hline
$M_X\le M_D$ & $a$ & 9.5 & 8.8 & 8.2 & 7.7 & 7.3 \\
 & Functional Form & 1.4\% & 1.1\% & 0.8\% & 0.5\% & 0.4\% \\
\hline
$M_X\le 1.7$\,GeV & $a$ & 12.5 & 11.5 & 10.5 & 9.7 & 8.9 \\
 & Functional Form & 2.9\% & 2.6\% & 2.2\% & 1.9\% & 1.6\% \\
\hline
$M_X\le 1.7$\,GeV & $a$ & 10.3 &  9.8 & 9.3 & 9.0 & 8.7 \\
$q^2\ge 8$\,GeV$^2$ & Functional Form & 2.0\% & 1.7\% & 1.5\% & 1.4\% & 1.4\%
 \\
\hline
$q^2\ge(M_B-M_D)^2$ & $a$ & 11.4 & 11.1 & 10.9 & 10.8 & 10.6 \\
 & Functional Form & 5.0\% & 4.4\% & 4.0\% & 3.6\% & 3.2\% \\
\hline
$P_+\le M_D^2/M_B$ & $a$ & 16.7 & 15.0 & 13.6 & 12.2 & 11.1 \\
 & Functional Form & 5.3\% & 4.8\% & 4.4\% & 4.0\% & 3.6\% \\
\hline
$E_l\ge 2.2$\,GeV & $a$ & 22.6 & 21.0 & 19.7 & 18.5 & 17.4 \\
 & Functional Form & 16.2\% & 13.1\% & 11.0\% & 9.3\% & 7.9\% \\
\hline
\end{tabular}
\end{center}
\end{table}

To explore the dependence on $m_b$ we define the exponent
\begin{equation}\label{eq:postulate}
   a(m_b)\equiv \frac{d\ln\Gamma_u}{d\ln m_b}
   = \left( \frac{\triangle\Gamma_u}{\Gamma_u} \right) /
   \left( \frac{\triangle m_b}{m_b} \right) ,
\end{equation}
which means that $\Gamma_u\sim (m_b)^a$. Table \ref{tab:mbdep} shows
the values of this exponent over a wide range of values of $m_b$ 
for a variety of experimental cuts. 
To estimate the sensitivity to the functional form we scan over a large set of
models for the subleading shape functions, and we also study the difference
between the results obtained using the exponential or the 
gaussian ansatz for the leading shape function. The corresponding variations
are added in quadrature and given as a relative change in the 
corresponding partial decay rates (labeled ``Functional Form'').
In all cases, $\mu_\pi^2=0.2$\,GeV$^2$ is kept fixed. 
Because we restrict ourselves to only two functional forms for the leading
shape function in this study, the resulting sensitivities should be 
interpreted with caution. 

The entries in the table are listed in roughly the order of increasing 
sensitivity to $m_b$ and to the functional form of the shape functions, with 
the hadronic invariant mass cut showing the least sensitivity and the 
lepton energy cut exhibiting the largest one. To some extent this reflects 
the different efficiencies (or ``inclusiveness'') of the various cuts.
It is reassuring that 
$a\approx 10$ for the pure $q^2$ cut, in accordance with the findings of
\cite{Neubert:2000ch,Neubert:2001ib}. Perhaps somewhat surprisingly, 
for this cut a substantial sensitivity to shape-function effects remains 
even for fixed $m_b$ and $\mu_\pi^2$. It is well known that the partial rate
with a cut $q^2\ge(M_B-M_D)^2$ can be calculated using a local OPE in powers
of $\Lambda_{\rm QCD}/m_c$ \cite{Neubert:2000ch,Bauer:2000xf}, thereby 
avoiding the notion of shape-function sensitivity. 
Differences between the functional forms of the shape functions 
in our approach correspond to effects 
that are formally of order $1/m_c^3$ and higher. It is not unreasonable that 
these effects should be of order 3--5\%.

We also checked that for much more relaxed cuts the value of $a(m_b)$
tends to $4.8$, as stated in (\ref{totrate}). For example, for a cut 
$P_+\le\Delta_P$ we find (with $m_b=4.61$\,GeV and $\mu_\pi^2=0.2$\,GeV$^2$):

\begin{center}
\begin{tabular}{|c|rrrrrrrr|} 
\hline
$\Delta_P$ [GeV] & 0.6 & 0.8 & 1.0 & 1.2 & 1.6 & 2.0 & 3.0 & $M_B$ \\ 
\hline
$a$ & 15.4 & 9.8 & 7.0 & 5.8 & 5.1 & 5.0 & 4.9 & 4.8 \\
\hline
\end{tabular}
\end{center}

\section{Conclusions}
\label{sec:concl}

A high-precision measurement of the parameters of the unitarity triangle 
is an ongoing quest, which
necessitates the close cooperation of theory and experiment. The
determination of $|V_{ub}|$ from inclusive 
$\bar B\to X_u\,l^-\bar\nu$ decay requires the measurement of partial decay
rates with kinematical cuts that eliminate the large background
from $\bar B\to X_c\,l^-\bar\nu$ decay, as well as theoretical
predictions for such quantities. To this end, it is desirable to have a
theoretical description of the triple differential decay rate, which can
be used for predicting arbitrary partial rates obtained after integrating 
over certain regions of phase space. One problem in providing such a 
description is that the power-counting rules of the heavy-quark expansion 
are different in different kinematical domains. In this paper we have 
overcome this difficulty. 

In the shape-function region, our results are in agreement with
QCD factorization theorems, and perturbative effects have been
separated from non-perturbative shape functions. When the allowed
phase space extends over a large domain, our results smoothly
reduce to the expressions obtained from the local operator product 
expansion. We have presented a formalism in which event distributions and
partial decay rates are expressed without explicit reference to
partonic quantities such as the $b$-quark mass. The sensitivity to such
hadronic parameters enters indirectly, via the moments of shape functions.
The most important non-perturbative object, namely the leading-order shape
function, can be extracted from the photon spectrum in 
$\bar B\to X_s\gamma$ decay. 
This is analogous to extractions of parton distribution 
functions from fits to data on deep inelastic scattering.
In this way, the dominant uncertainty from our ignorance about bound-state 
effects in the $B$ meson is turned into an experimental 
uncertainty, which will reduce with increasing accuracy of the experimental 
data on the photon spectrum. Residual hadronic uncertainties are power
suppressed in the heavy-quark expansion.

One goal of this paper was to present a detailed framework
in which this program can be carried out. We have given formulae
that can be readily used for the construction of an
event generator, as well as to estimate the remaining
theoretical uncertainties in a robust and automated fashion.

In practice the leading shape function needs to be parameterized. We have 
suggested three different functional forms, which can be used to fit the data 
of the $\bar B\to X_s\gamma$ photon spectrum. Once the data is accurately 
described by a choice of the shape functions, this function can be used in
the predictions for partial $\bar B\to X_u\,l^-\bar\nu$ rates and
spectra. Subleading shape functions give rise to
theoretical uncertainties starting at the level of $1/m_b$ power corrections. 
We have estimated these uncertainties using a large set of models, each of 
which obeys the known tree-level moment relations, but which are very
different in their functional form. A second error estimate is
determined by the residual renormalization-scale dependence. We also
considered uncertainties from weak annihilation effects, which in principle 
can be avoided by cutting away the region of phase space in which they
contribute. We have suggested a cut on high leptonic invariant mass,
which accomplishes just that.

The second half of this paper contains detailed 
numerical predictions for a variety of partial $\bar B\to X_u\,l^-\bar\nu$ 
decay rates with different kinematical cuts, 
including cuts on the charged-lepton energy (both
in the rest frame of the $B$ meson and of the $\Upsilon(4S)$ resonance),
on the hadronic quantity $P_+=E_X-|\vec{P}_X|$, on $M_X$, on $q^2$, and on 
various combinations of these variables. Along with our predictions 
for the rates we have presented a complete analysis of theoretical
uncertainties. Once the data on
the $\bar B\to X_s\gamma$ photon spectrum are sufficiently precise to
accurately determine the leading-order shape function, a determination of
$|V_{ub}|$ with theoretical uncertainties at the 5--10\% level now seems 
feasible.

\vspace{0.3cm}
\noindent 
{\em Acknowledgments:} 
We are grateful to many of our colleagues in both experiment and theory. We 
thank Christian Bauer, 
Ilija Bizjak, Riccardo Faccini, Lawrence Gibbons, Vladimir Golubev,
Kay Kinoshita, Robert Kowalewski, Seung Lee, Zoltan Ligeti, Antonio Limosani,
Francesca Di~Lodovico, Vera Luth, Thomas Meyer, Masahiro Morii,
Franz Muheim, and Tadao Nozaki for valuable
discussions. We would like to thank the Institute for Advanced Study,
Princeton, NJ, where part of this work was done, for their
hospitality. The work of B.O.L. was supported in part by funds
provided by the U.S.~Department of Energy (D.O.E.) under cooperative
research agreement DE-FC02-94ER40818. The research of M.N. and
G.P. was supported by the National Science Foundation under Grant
PHY-0355005.

\newpage
\begin{appendix}

\section{Perturbative Expressions}
\label{apx:Sudakovs}

\subsection{Anomalous dimensions}

Here we list the known perturbative expansions of the $\beta$-function
and relevant anomalous dimensions. We work in the $\overline{\rm MS}$
scheme and define
\begin{eqnarray}
   \beta(\alpha_s)
   &=& \frac{d\alpha_s(\mu)}{d\ln \mu} 
    = -2\alpha_s \sum\limits_{n=0}^\infty \beta_n 
    \left( \frac{\alpha_s}{4\pi} \right)^{n+1} , \nonumber\\
   \Gamma_{\rm cusp}(\alpha_s) &=& \sum\limits_{n=0}^\infty \Gamma_n 
    \left( \frac{\alpha_s}{4\pi} \right)^{n+1} , \qquad
   \gamma'(\alpha_s) = \sum\limits_{n=0}^\infty \gamma'_n 
    \left( \frac{\alpha_s}{4\pi} \right)^{n+1} ,
\end{eqnarray}
as the expansion coefficients for the $\beta$-function, the leading-order 
SCET current anomalous dimension, and the cusp anomalous
dimension. To three-loop order, the $\beta$-function reads \cite{Tarasov:au} 
\begin{eqnarray}
   \beta_0 &=& \frac{11}{3}\,C_A - \frac23\,n_f \,, \qquad
    \beta_1 = \frac{34}{3}\,C_A^2 - \frac{10}{3}\,C_A\, n_f - 2C_F\,n_f \,, 
    \nonumber \\
   \beta_2 &=& \frac{2857}{54}\,C_A^3
    + \left( C_F^2 - \frac{205}{18}\,C_F C_A
    - \frac{1415}{54}\,C_A^2 \right) n_f
    + \left( \frac{11}{9}\,C_F + \frac{79}{54}\,C_A \right) n_f^2 \,,
\end{eqnarray}
where $n_f=4$ is the number of light flavors, $C_A=3$ and $C_F=4/3$. The 
three-loop expression for the cusp anomalous dimension has
recently been obtained in \cite{Moch:2004pa}. The coefficients read
\begin{eqnarray}
   \Gamma_0 &=& 4C_F \,, \qquad
    \Gamma_1 = 8C_F \left[ \left( \frac{67}{18} - \frac{\pi^2}{6} \right) C_A 
    - \frac59\,n_f \right] , \nonumber\\
   \Gamma_2 &=& 16C_F \bigg[ \left(\frac{245}{24} - \frac{67\pi^2}{54}
    + \frac{11\pi^4}{180} + \frac{11}{6}\,\zeta_3 \right) C_A^2
    + \left( - \frac{209}{108} + \frac{5\pi^2}{27} - \frac73\,\zeta_3
    \right) C_A\,n_f \nonumber\\
   &&\mbox{}+ \left( - \frac{55}{24} + 2\zeta_3 \right) C_F\,n_f
    - \frac{1}{27}\,n_f^2 \bigg] \,.
\end{eqnarray}
The SCET anomalous dimension $\gamma$ is explicitly known only to
one-loop order. However, the two-loop coefficient can be extracted by
noting that $\gamma$ is related to the axial-gauge anomalous dimension
in deep inelastic scattering \cite{Neubert:2004dd}. The result is
\begin{eqnarray}
   \gamma_0' &=& -5 C_F \,, \\
   \gamma_1' &=& -8 C_F \bigg[ \left(\frac{3}{16} - \frac{\pi^2}{4}
    + 3\zeta_3 \right) C_F
    + \left( \frac{1549}{432} + \frac{7\pi^2}{48} - \frac{11}{4}\,\zeta_3
    \right)\,C_A
    - \left( \frac{125}{216} + \frac{\pi^2}{24} \right) n_f \bigg] \,.
    \nonumber
\end{eqnarray}

\subsection{Evolution factor}

The exact expression for the evolution factor reads
\begin{equation}\label{eq:U1}
   \ln U(\mu_h,\mu_i) = 2 S_\Gamma(\mu_h,\mu_i)  
   - 2a_\Gamma(\mu_h,\mu_i) \,\ln \frac{m_b}{\mu_h} 
   - 2a_{\gamma'}(\mu_h,\mu_i) \,,
\end{equation}
where the functions of the right-hand side are solutions to the 
renormalization-group equations
\begin{eqnarray}\label{eq:aGammaETC}
   \frac{d}{d\ln\mu} S_\Gamma(\nu,\mu)
   &=& - \Gamma_{\rm cusp}(\alpha_s(\mu))\,\ln\frac{\mu}{\nu} \,, \nonumber\\
   \frac{d}{d\ln\mu} a_\Gamma(\nu,\mu)
   &=& - \Gamma_{\rm cusp}(\alpha_s(\mu)) \,, \qquad
   \frac{d}{d\ln\mu} a_{\gamma'}(\nu,\mu) = - \gamma'(\alpha_s(\mu)) \,,
\end{eqnarray}
with boundary conditions $S(\nu,\mu)=0$ etc.\ at $\mu=\nu$.
These equations can be integrated using that 
$d/d\ln\mu=\beta(\alpha_s)\,d/d\alpha_s$. The solutions are
\begin{equation}
   S_\Gamma(\nu,\mu)
   = -\int\limits_{\alpha_s(\nu)}^{\alpha_s(\mu)}\!\!d\alpha\,
    \frac{\Gamma_{\rm cusp}(\alpha)}{\beta(\alpha)} 
    \int\limits_{\alpha_s(\nu)}^{\alpha}\frac{d\alpha'}{\beta(\alpha')}
    \,, \qquad
   a_\Gamma(\nu,\mu)
   = -\int\limits_{\alpha_s(\nu)}^{\alpha_s(\mu)}\!\!d\alpha\,
    \frac{\Gamma_{\rm cusp}(\alpha)}{\beta(\alpha)} \,,
\end{equation}
and similarly for $a_{\gamma'}$. 

Next, we give explicit results for the Sudakov exponent $S_\Gamma$ and
the functions $a_\Gamma$ and $a_\gamma$ in (\ref{eq:U1}) at next-to-leading 
order in renormalization-group improved perturbation theory. We obtain
\begin{equation}
   a_\Gamma(\nu,\mu) = \frac{\Gamma_0}{2\beta_0}
   \left[ \ln\frac{\alpha_s(\mu)}{\alpha_s(\nu)}
   + \left( \frac{\Gamma_1}{\Gamma_0} - \frac{\beta_1}{\beta_0} \right)
   \frac{\alpha_s(\mu)-\alpha_s(\nu)}{4\pi} + \dots \right] ,
\end{equation}
and similarly for $a_\gamma$. The next-to-leading order expressions for the 
Sudakov exponent $S_\Gamma$ contains the three-loop coefficients $\beta_2$ 
and $\Gamma_2$. With $r=\alpha_s(\mu)/\alpha_s(\nu)$, it reads
\begin{eqnarray}
   S_\Gamma(\nu,\mu)
   &=& \frac{\Gamma_0}{4\beta_0^2}\,\Bigg\{
    \frac{4\pi}{\alpha_s(\nu)} \left( 1-\frac1r-\ln r \right)
    + \left( \frac{\Gamma_1}{\Gamma_0} - \frac{\beta_1}{\beta_0} \right)
    (1-r+\ln r) + \frac{\beta_1}{2\beta_0} \ln^2 r \nonumber\\
  &&\mbox{}+ \frac{\alpha_s(\nu)}{4\pi} \Bigg[ \left( 
   \frac{\beta_1\Gamma_1}{\beta_0\Gamma_0}-\frac{\beta_2}{\beta_0}
   \right) (1-r+r\ln r) + \left( 
   \frac{\beta_1^2}{\beta_0^2}-\frac{\beta_2}{\beta_0}\right)
   (1-r) \ln r \nonumber\\
  &&\hspace{1cm}\mbox{}- \left( \frac{\beta_1^2}{\beta_0^2}
   - \frac{\beta_2}{\beta_0}-\frac{\beta_1\Gamma_1}{\beta_0\Gamma_0}
   + \frac{\Gamma_2}{\Gamma_0} \right) \frac{(1-r)^2}{2}
   \Bigg] + \dots \Bigg\} \,.
\end{eqnarray}
The next-to-leading-logarithmic evolution factor $U(\mu_h,\mu_i)$
can be obtained by combining the above expressions according to (\ref{eq:U1}) 
and expanding out terms of order $\alpha_s$.

\section{Partially integrated decay rates}
\label{appendix:B}

With the exception of the combined cut on the lepton energy $E_l$ and
the hadronic quantity $s_H^{\rm max}$ studied in
Section~\ref{sec:babarcut}, all other partial rates investigated in our 
analysis can be derived by first integrating the triple differential decay 
rate (\ref{eq:tripleRate}) over the lepton energy $E_l\ge E_0$ 
and $P_-\le P_-^{\rm max}$ analytically, where the quantity 
$P_-^{\rm max}$ (and in principle even $E_0$) may depend on the value of 
$P_+$. The remaining integration over $P_+$ is then performed numerically. 
In such a situation, we need to evaluate the partially integrated decay rate
\begin{equation}\label{PSints}
   \frac{d\Gamma_u}{dP_+}
   = \int_{P_+}^{P_-^{\rm max}}\!dP_-
   \int_{P_+}^{{\rm min}(P_-,M_B-2E_0)}\!dP_l\,
   \frac{d^3\Gamma_u}{dP_l\,dP_-\,dP_+} \,.
\end{equation}
Changing variables from $P_-$ to
$y$ defined in (\ref{ydef}), the constraint $P_-\le P_-^{\rm max}$
translates into the integration domain $0\le y\le y_{\rm max}$, where in 
analogy to (\ref{ydef}) we define
\begin{equation}
   y_{\rm max} = \frac{P_-^{\rm max}-P_+}{M_B-P_+} \,, \qquad
   y_0 = \frac{P_l^{\rm max}-P_+}{M_B-P_+} = 1 - \frac{2 E_0}{M_B-P_+} \,.
\end{equation}
From the phase-space relation (\ref{eq:simplePhaseSpace}) it follows
that a cut on the lepton energy has no effect if $y_0\ge y_{\rm max}$. 
The result of performing the integrations in (\ref{PSints}) can be written as
\begin{equation}
   \frac{d\Gamma_u(y_{\rm max},y_0)}{dP_+}
   = \left\{ \begin{array}{lll}
   \Gamma_u^A(y_{\rm max}) &;& \quad y_{\rm max} \le y_0 \,, \\
   \Gamma_u^A(y_0) + \Gamma_u^B &;& \quad y_{\rm max} > y_0 \,,
   \end{array} \right.
\end{equation}
where 
\begin{eqnarray}
   \Gamma_u^A(y_i)
   &=& \frac{G_F^2|V_{ub}|^2}{96\pi^3}\,(M_B-P_+)^5\,U(\mu_h,\mu_i)
    \int_0^{y_i}\!dy\,y^{2-2a_\Gamma} \left[ (3-2y)\,\F_1
    + 6(1-y)\,\F_2 + y\,\F_3 \right] , \nonumber\\
   \Gamma_u^B
   &=& \frac{G_F^2|V_{ub}|^2}{96\pi^3}\,(M_B-P_+)^5\,U(\mu_h,\mu_i)
    \int_{y_0}^{y_{\rm max}}\!dy\,y^{-2a_\Gamma} y_0 \nonumber\\
   &&\times \left[ \left( 6y(1+y_0) - 6y^2
    - y_0(3+2y_0) \right) \F_1
    + 6y(1-y)\,\F_2
    + y_0(3y-2y_0)\,\F_3 \right] .
\end{eqnarray}
When the kinematical power corrections in (\ref{eq:fullDFN}) are 
expanded as in (\ref{eq:kinNLO}) and (\ref{eq:kinNNLO}), the resulting
integrals over $y$ can be expressed in terms of the master functions
$I_n(b,z)$ given in eq.~(86) of \cite{Bosch:2004th}. The resulting expressions 
are used to obtain the numbers in the various tables in 
Section~\ref{sec:analysio}.

We now list the values of $y_0$ and $y_{\rm max}$ for the different cuts 
studied in Section~\ref{sec:analysio}. Whenever a cut $E_l\ge E_0$ on the 
charged-lepton energy is applied, we have
\begin{equation}
    y_0 = 1 - \frac{2E_0}{M_B-P_+} \,.
\end{equation}
For an additional cut $P_+\le\Delta_P$, we have $y_{\rm max}=1$ and 
$0\le P_+\le\mbox{min}(\Delta_P,M_B-2E_0)$. 
For a cut on hadronic invariant mass, $M_X\le M_0$, we have
\begin{equation}
   y_{\rm max} = \frac{\mbox{min}(M_B,M_0^2/P_+) - P_+}{M_B - P_+}
\end{equation}
and $0\le P_+\le\mbox{min}(M_0,M_B-2E_0)$.
For a cut on leptonic invariant mass, $q^2\ge q_0^2$, we have
\begin{equation}
   y_{\rm max} = 1 - \frac{q_0^2}{(M_B - P_+)^2}
\end{equation}
and $0\le P_+\le\mbox{min}(M_B-q_0,M_B-2E_0)$. 
Finally, for the combined $M_X$--$q^2$ cut we take the minimum of the previous 
two $y_{\rm max}$ values.

\end{appendix}

\end{document}